\documentclass[traditabstract]{aa} 
\usepackage{graphicx}
\usepackage{txfonts}
\usepackage{natbib}
\bibpunct{(}{)}{;}{a}{,}

\newcommand\lesim{\lower.5ex\hbox{$\; \buildrel < \over \sim \;$}}
\newcommand\gesim{\lower.5ex\hbox{$\; \buildrel > \over \sim \;$}}
\newcommand\HI{H{\small I}}
\newcommand\kms{km s$^{-1}$}

\newcommand\dg{$^{\circ }$}
\newcommand\sn{S/N}

\begin{document}
   \title{The dark matter halo shape of edge-on disk galaxies}

\subtitle{II.  Modelling the HI observations: methods}

\titlerunning{The dark matter halo shape of edge-on disk galaxies II}
   \author{J.C. O'Brien
          \inst{1},
K.C. Freeman\inst{1}
          \and
          P.C. van der Kruit\inst{2}}

   \institute{Research School of Astronomy and Astrophysics, 
Australian National
University, Mount Stromlo Observatory, Cotter Road, ACT 2611, Australia\\
              \email{jesscobrien@gmail.com; kcf@mso.anu.edu.au$^\star$}
         \and
             Kapteyn Astronomical Institute, University of Groningen, 
P.O. Box 800,
9700 AV Groningen, the Netherlands\\
             \email{vdkruit@astro.rug.nl}\thanks{For 
correspondence contact Ken Freeman or 
Piet van der Kruit}
             }
\authorrunning{J.C. O'Brien et al.}

   \date{Received Xxxxxxx 00, 2010; accepted Xxxxxxx 00, 2010}

 
  \abstract
   {This is the second paper of a series in which we attempt to put
constraints on the flattening of dark halos in disk galaxies. For 
this purpose, we
observe the \HI\ in edge-on galaxies, where it is in principle
possible to measure the force field in the halo vertically and radially
from gas layer flaring and rotation curve decomposition respectively.  
To calculate the force fields, we need to analyse the observed XV diagrams 
to accurately measure all three functions that describe the
planar kinematics and distribution of a galaxy: the radial \HI\ surface
density, the rotation curve and the \HI\ velocity dispersion. 
In this  paper, we discuss the improvements and limitations of the methods 
previously used to measure these \HI\ properties. We extend the 
constant velocity dispersion method to include determination of the \HI\
velocity dispersion as a function of galactocentric radius and perform
extensive tests on the quality of the fits.

We will apply this `radial decomposition XV modelling method' to our 
\HI\ observations of 8 \HI\ rich,
late-type, edge-on galaxies in the third paper of this series.  }
   {}


   \maketitle
%

\section{Introduction}
In paper I \citep{ofk2008}
in this series we presented \HI\ observations of a sample of 8
edge-on, \HI\ rich, late-type galaxies. The aim of the project has been
described there in detail. Briefly, we attempt to put 
constraints on the flattening of dark halos around disk galaxies by
measuring the force field of the halo vertically from the flaring of
the \HI\ layer and radially from rotation curve decomposition. For the
vertical force field we need to determine in these galaxies both the
velocity dispersion of the \HI\ gas (preferably as a function of
height from the central plane of the disk) and the thickness of the \HI\
layer, all of this as a function of galactocentric radius. In addition
we also need to extract information on the rotation curve of the galaxy
and the deprojected
\HI\ surface density, as a function of galactocentric radius. In this
second paper in the series 
we will review earlier determinations of these properties and then describe
the methods we have developed to analyse our sample.


\begin{table*}[t] 
\caption[Summary of previous measurements of gas velocity dispersion 
in face-on galaxies]
{Summary of   previous observations of gas velocity dispersion in 
face-on galaxies.}
\label{tab:ch4-vel-disp-meas-lowi}
\begin{footnotesize}
\begin{tabular}{lcccccccccl}
\hline
Galaxies & i & Type & H{\scriptsize I} Velocity & 
\multicolumn{2}{c}{Resolution} &
\multicolumn{3}{c}{H{\scriptsize I} radius} & Dist. & Reference \\
\cline{5-9}
& & & dispersion & Spectral &
Spatial & & & & & \\
\cline{5-6}
& \dg & & & \kms & kpc & kpc & $R_{25}$ & $\theta_{maj}$ & Mpc & \\
\hline\hline
NGC3938 & $10\pm5$ & ScI & constant$^{\rm h}$ & 8.2$^{\rm a}$ &
$1.2 \times 1.7$ & 13.1 & 1.7
& 7.5 & 10 & \citet{vdks1982c} \\
NGC628 & $5\pm5$$^{\rm e}$ & ScI & falling$^{\rm h}$ & 8.2$^{\rm a}$ &
 $2.9 \times 2.9$ & 40.7 & 2.8 & 14.0 &
10 & \citet{svdk1984} \\
NGC1058 & $8\pm2$$^{\rm f}$ & Sc{\sc II-III} & constant$^{\rm h}$ &
8.2$^{\rm a}$ &
$2.2 \times 2.2$$^{\rm c}$ & 18.9 & 4.3 & 8.7 & 10$^{\rm c}$ &
\citet{vdks1984} \\
NGC1058 & $8\pm2$$^{\rm f}$ & Sc{\sc II-III} & falling$^{\rm h}$ &
2.58$^{\rm a}$ &
$2.1 \times 2.0$$^{\rm d}$ & 14.1 & 3.2 & 6.7 & 10$^{\rm d}$ &
\citet{dhh1990} \\
NGC6946 & 34 & Scd/Sc{\sc I} & falling$^{\rm h}$ & 8.25$^{\rm a}$ &
$0.7 \times 0.8$ & 17.2 & 2.0 & 21.2 & $5.9$$^{\rm b}$ &
\citet{bv1992} \\
NGC5474$^{\rm g}$ & $21^{+4}_{-6}$ & ScdIV & falling$^{\rm h}$ &
5.2$^{\rm a}$ & $1.2
\times 1.1$ & 10.5 & 4.6 & 8.6 & 7 &
\citet{rdh1994} \\
NGC1058 & $8\pm2$$^{\mathrm f}$ & Sc{\sc II-III} & falling$^{\rm h}$ &
2.58$^{\rm a}$ &
$1.5 \times 1.5$ & 21.8 & 5.0 & 15.0 & 10 &
\citet{pr2006} \\
\hline
\end{tabular}
\begin{list}{}{}
\item[$^{\mathrm{a}}$] Hanning smoothed.
\item[$^{\mathrm{b}}$] Distance shown is mean distance calculated 
from brightest
    stars in NGC6946 group.
\item[$^{\mathrm{c}}$] To facilitate comparison with the other observations for
    NGC1058, we have used the adopted distance
    of $10$ Mpc used by \citet{pr2006}, rather than
    the \citet{vdks1984} preferred value of $9$ Mpc.
\item[$^{\mathrm{d}}$] To facilitate comparison with the other observations for
    NGC1058, we have used the adopted distance
    of $10$ Mpc used by \citet{pr2006}, rather than
    the \citet{dhh1990} preferred value of $10.2-14.5$ Mpc.
\item[$^{\mathrm{e}}$] Warps from $\approx5^{\circ}$ within the optical disk to
    $i\approx0^{\circ}$, then back to $i\approx10^{\circ}$ at the 
H{\scriptsize I}
    edge \citep{svdk1984}.
\item[$^{\mathrm{f}}$] Warps from $\approx8\pm2^{\circ}$ within the optical 
disk to
    exactly face-on at the H{\scriptsize I} edge \citep{vdks1984}
\item[$^{\mathrm{g}}$] Classified as peculiar, and known to be interacting 
with its
    much larger neighbour M101 which is only $90$ kpc distance
    assuming a distance of $7$ Mpc \citep{rdh1994}.
\item[$^{\mathrm{h}}$] Direct measurement of the velocity dispersion of the
    broadening of the vertical velocity dispersion from a near face-on
    disk. Measurements corrected for instrumental broadening and
    inclination, but not beam-smearing.
\end{list}
\end{footnotesize}
\end{table*}

To meet our goals we need to measure the vertical structure of galaxies
out to the lowest surface densities possible and use galaxies with
relatively large fractions of their mass in their halos, i.e. high total 
mass-to-light ratios galaxies. For that purpose
we defined a sample of nearby, \HI\ rich, late-type edge-on galaxies.
From paper I we recall that for a vertically isothermal gas sheet 
with a vertically Gaussian density distribution, the gradient of 
the total force in the vertical direction can be 
written as

\begin{equation}
\label{eq:ch1-hydro}
\frac{\partial K_z}{\partial z} = -
\frac{\sigma_{v,g}^2}{({\rm FWHM}_{z,g}/2.35)^2},
\end{equation}
where $\sigma_{v,g}$ is the vertical velocity dispersion of the gas and
FWHM$_{z,g}$ the gas layer thickness. 

The vertical force field comes from the gas itself, plus the dark halo
and the stars of the disk. In regions where the halo is a significant
contributor to $K_z$, the \HI\ flaring will be smaller for the more
flattened halos with a smaller axis ratio $q=c/a$ (the ratio of the halo 
polar axis $c$ to its major axis in the galactic plane $a$). 

In the present paper 
we will first present a new method to accurately determine the
rotation curve, deprojected \HI\ surface density and \HI\ velocity
dispersion at all radii in a edge-on gas disk. The motivation for 
this measurement method came from the strong dependence in
Eqn.~(2) in paper I of the derived dark halo flattening
$q$ on the measured \HI\ velocity dispersion and vertical \HI\
gas disk flaring. The
superposition of velocity profiles from many radii in each sightline
through an edge-on \HI\ disk tends to cause an overestimate of the
velocity dispersion with most measurement methods, as explained below.
Measuring the radial flaring profile requires a model of the galaxy
rotation and face-on surface density; this also necessitates high
accuracy rotation curve measurement and \HI\ surface density
deprojection.

The \HI\ velocity dispersion has not previously been measured
systematically in edge-on galaxies. Most measurements of the \HI\
velocity dispersion have been conducted on face-on galaxies, and these
measurements were largely conducted 15-20 years ago with relatively
low spatial resolution FWHM$_{\theta}$= 1-3 kpc. In
Table~\ref{tab:ch4-vel-disp-meas-lowi} and
\ref{tab:ch4-vel-disp-meas-highi}, we show the resolution of each
observation used to measure the velocity dispersion of near face-on
and near-edge-on galaxies, respectively. The radial structure of the
\HI\ velocity dispersion has been measured in only 6 galaxies -- 5 from
near face-on galaxies and one from an edge-on galaxy. The 5 face-on
galaxies were observed at $\sim$1 kpc or larger resolution. The
uncertain inclination of low resolution face-on \HI\ observations can
cause additional broadening, due to projection of the gas rotation
along the line-of-sight. Galaxies inclined slightly away from face-on
will suffer beam smearing of the gradients in the projected rotation
as well as beam smearing of the gradients of the intrinsic velocity
dispersion.

\begin{table*}[t] 
\caption[Summary of previous measurements of gas velocity 
dispersion in highly-inclined 
galaxies]{Summary of 
  previous observations of gas velocity dispersion of highly inclined galaxies.}
\label{tab:ch4-vel-disp-meas-highi}
\begin{footnotesize}
\begin{tabular}{lcccccccccc}
\hline
Galaxies & i & Type & H{\scriptsize I} Velocity & 
\multicolumn{2}{c}{Resolution} &
\multicolumn{3}{c}{H{\scriptsize I} radius} & Distance & Ref. \\
\cline{5-9}
& & & dispersion & Spectral &
Spatial &  &  & & & \\
\cline{5-6}
& \dg & & & \kms & kpc & $R_{25}$ & kpc & $\theta_{maj}$ & Mpc &  \\
\hline\hline
NGC4244 & $84.5$$^{\rm c}$ & Scd & constant$^{\rm d}$ &
5.2$^{\rm a}$ & 0.17 & 630 & 11.0 & 2.0 & 3.6 &
\citet{olling1996a} \\
ESO142-G24  & 90$^{\rm b}$    & Scd    &
constant$^{\rm e}$ & 6.6$^{\rm a}$   & 3.8
& 19.0    & 1.6      & 5.0    & 26.6 &
\citet{kvdkdb2004} \\
ESO157-G18  & 90$^{\rm b}$    & Scd    &
constant$^{\rm e}$ & 6.6$^{\rm a}$    & 2.6     &
11.1    & 1.3      & 4.3    & 18.1 &
\citet{kvdkdb2004} \\
ESO201-G22  & 90$^{\rm b}$    & Sc    &
constant$^{\rm e}$ & 6.6$^{\rm a}$     & 9.0     &
34.1    & 1.5      & 3.8    & 55.4 &
\citet{kvdkdb2004} \\
ESO240-G11  & 90$^{\rm b}$    & Sc   &
constant$^{\rm e}$ & 6.6$^{\rm a}$     & 1.7     &
41.5    & 1.4     & 24.2    & 38.1 &
\citet{kvdkdb2004} \\
ESO263-G15  & 90$^{\rm b}$    & Sc    &
constant$^{\rm e}$ & 6.6$^{\rm a}$     & 5.1     &
24.8    & 1.5      & 4.8    & 35.2 &
\citet{kvdkdb2004} \\
ESO269-G15  & 90$^{\rm b}$    & Sc    &
constant$^{\rm e}$ & 6.6$^{\rm a}$     & 7.5     &
19.9    & 1.0      & 2.7    & 46.7 &
\citet{kvdkdb2004} \\
ESO321-G10  & 90$^{\rm b}$    & Sa    &
constant$^{\rm e}$ & 6.6$^{\rm a}$     & 7.3       &
14.5    & 1.2      & 2.0    & 41.0 &
\citet{kvdkdb2004} \\
ESO416-G25  & 90$^{\rm b}$    & Sb    &
constant$^{\rm e}$ & 6.6$^{\rm a}$     & 17.7     &
36.7    & 1.6      & 2.1    & 67.0 &
\citet{kvdkdb2004} \\
ESO435-G14  & 90$^{\rm b}$    & Sc     &
constant$^{\rm e}$ & 6.6$^{\rm a}$     & 7.7     &
18.9    & 1.5      & 2.5    & 35.2 &
\citet{kvdkdb2004} \\
ESO435-G25  & 90$^{\rm b}$    & Sc   & constant$^{\rm e}$ &
6.6$^{\rm a}$     & 2.2     & 36.1    & 1.2     & 16.3    & 32.4 &
\citet{kvdkdb2004} \\
ESO435-G50  & 90$^{\rm b}$    & Sc    & constant$^{\rm e}$ &
6.6$^{\rm a}$     & 7.1       & 13.1    & 1.4      & 1.9    & 35.2 &
\citet{kvdkdb2004} \\
ESO446-G18  & 90$^{\rm b}$    & Sb    & constant$^{\rm e}$ &
6.66$^{\rm a}$     & 12.2       & 31.1    & 1.3      & 2.5    & 64.1 &
\citet{kvdkdb2004} \\
ESO564-G27  & 90$^{\rm b}$    & Sc   & constant$^{\rm e}$ &
6.66$^{\rm a}$     & 10.2       & 28.7    & 1.6      & 2.8    & 29.5 &
\citet{kvdkdb2004} \\
NGC5170     & 90$^{\rm b}$    & Sc   & constant$^{\rm e}$ &
6.66$^{\rm a}$     & 7.4        & 31.9    & 1.3      & 4.3    & 20.9 &
\citet{kvdkdb2004} \\
\hline
\end{tabular}
\begin{list}{}{}
\item[$^{\mathrm{a}}$] Hanning smoothed.
\item[$^{\mathrm{b}}$] From LEDA database.
\item[$^{\mathrm{c}}$] The unwarped inner disk has an inclination of
    $i=84.5^{\circ}$, outside of which there is a small warp to
    $i=82.5\pm1^{\circ}$, before warping back to the plane of the
    inner disk at large radii \citep{olling1996a}.
\item[$^{\mathrm{d}}$] Direct measurement of the velocity dispersion of the
    broadening of the extreme velocity envelope of the XV map of an
    edge-on galaxy. Measurements corrected for instrumental
    broadening, projection and beam-smearing, via Olling's method (see
    Sect.~\ref{sec:olling-RC-disp-meth}).
\item[$^{\mathrm{e}}$] Iterative envelope fitting of the XV map of 
an high-inclination galaxy
    integrated over the galaxy minor axis \citep{kvdkdb2004}. This
    method was unable to measure radial structure of the velocity
    dispersion, as the method did not include beam correction and the
    galaxies in the sample were observed with very large beams.
\end{list}
\end{footnotesize}
\end{table*}

\begin{table*}[t] 
  \caption[Summary of previous measurements of rotation curves in
  edge-on galaxies]{Summary of previous measurements of rotation
    curves in edge-on galaxies ($i > 80^{\circ}$).}
  \label{tab:ch4-RC-meas}
  \begin{footnotesize}
    \begin{tabular}{lp{4.0cm}ccccl}
      \hline
      Galaxies & Method & Line &
      \multicolumn{2}{c}{Resolution} & Distance & Reference \\
      \cline{4-5}
      & & & Spectral & Spatial & & \\
      & & & \kms & kpc & Mpc & \\
      \hline\hline
     UGC7321 & Peak Flux$^{\rm a}$ & H{\scriptsize I} & 5.2 & 0.8 & 10 &
      \protect\citet{um2003} \\
      \hline
      NGC891 & Envelope Tracing$^{\rm b}$ & H{\scriptsize I} & 27 & 1.7 &
      9.5$^{\rm f}$ &      \protect\citet{sa1979} \\
      NGC891$^{\ddagger}$ & Envelope Tracing$^{\rm b}$ & H{\scriptsize I} & 33 &
      0.7 & 9.5$^{\rm f}$ & \protect\citet{ssvdh1997} \\
      9 dwarfs & Envelope Tracing$^{\rm b,e}$ & H{\scriptsize I} & 4-8 & 0.5-2.8 &
      3.4-19 & \protect\citet{swaters1999} \\
      4 spirals & Envelope Tracing$^{\rm b}$ & HI & 20-41 & 0.9-2.3 &
      9.5-15.5 &
      \protect\citet{sofue1996} \\
      4 LSBs$^{\#}$ & Envelope Tracing$^{\rm b}$ & H{\scriptsize I} & 4-8 & 0.7-2.7 &
      5-19 & \protect\citet{dbb2002} \\
      \hline
      NGC891$^{\dagger}$ & Fixed $\sigma_v$ \& Parametric $v(R)$
      XV Modelling$^{\rm b,c}$ &H{\scriptsize I} & 27 & 1.7 &
      9.5 &
      \protect\citet{vdkruit1981} \\
      NGC5023 & Fixed $\sigma_v$ \& Parametric $v(R)$ XV
      Modelling$^{\rm b,c}$ & H{\scriptsize I} & 17
      & 0.8 & 7.9 & \protect\citet{bsvdk1986} \\
      \hline
      NGC3079 & Iterative Envelope Tracing$^{\rm b}$ & CO
      & 5.2 & 0.12 &
      15.5 & \protect\citet{ts2002} \\
      \hline
      14 spirals & Fixed $\sigma_v$ 1D Gaussian Envelope
     Fitting$^{\rm b}$ & H{\scriptsize I} &
      6.6 & 1.7-17.7 & 18-67 & \protect\citet{kvdkdb2004} \\
      \hline
      24 spirals & Fixed $\sigma_v$ 2D Gaussian Envelope
      Fitting$^{\rm b}$ & H{\scriptsize I} & 5.0-20.1 & 0.8-4.8 & 5-33 &
      \protect\citet{grsk2002} \\
      \hline
      8 spirals$^{\star}$ & Fixed $\sigma_v$ XV Modelling & H{\scriptsize I} &
      6.6-33.0 & 0.9-9.8
      & 9.5-54$^{\rm f}$ & \protect\citet{kvdk2004} \\
      \hline
      NGC4244 & Corrected Gaussian Envelope Fitting$^{\rm d}$ & H{\scriptsize I} &
      5.2 & 0.17 & 3.6 & \protect\citet{olling1996a} \\
      \hline
    \end{tabular}
\begin{list}{}{}
\item[$^{\mathrm{a}}$] Corrected for instrumental broadening \& gas velocity
        dispersion using the ``equivalent rectangular measure''
        \citep{sbs1966}.
\item[$^{\mathrm{b}}$] Corrected for instrumental broadening \& gas velocity
      dispersion by quadratic subtraction, 
see Eqn.~(\ref{eq:ch4-instr-broadening-corr}).
\item[$^{\mathrm{c}}$] Uses the standard \citet{fe1980} parametric rotation
        curve form, constant $\sigma_z$, and a 2D exponential surface
        density to model the XV diagram.
\item[$^{\mathrm{d}}$] Corrected for broadening \& beam-smearing (see Olling's
        method in Sect.~\ref{sec:olling-RC-disp-meth}).
\item[$^{\mathrm{e}}$] Asymmetric drift correction applied.
\item[$^{\mathrm{f}}$] Using distance measured by \citet{vdkruit1981} for 
NGC891.
\item[$^{\mathrm{\dagger}}$] Re-analysis of NGC891 data observed by 
\citet{sa1979}.
\item[$^{\mathrm{\ddagger}}$] New deeper WSRT NGC891 data -- $144$ hr.
\item[$^{\mathrm{\#}}$] Re-analysed subset of the \citet{swaters1999} dataset.
\item[$^{\mathrm{\star}}$] Includes re-analysed NGC891 data observed by 
\citet{ssvdh1997}.
\end{list}
  \end{footnotesize}
\end{table*}

From Table~\ref{tab:ch4-vel-disp-meas-lowi}, it is seen that only 3 of
the low-inclination velocity dispersion studies were undertaken with
good spectral resolution (channel FWHM$_v\lesim 5$ \kms). Of these
three high spectral resolution studies, only the recent measurements
of NGC1058 by \citet{pr2006} had sufficient spatial resolution to
sample the \HI\ radius by at least 10 synthesised beam FWHMs 
($\theta _{maj}$). The main
difference between the observations of \citet{pr2006} and the other
low inclination \HI\ velocity dispersion measurements are that the
observations by \citet{pr2006} were deeper ($23$ hr total integration
time on 3 VLA array configurations) and with good spectral resolution.
The two low \HI\ inclination velocity dispersion studies that found a
constant velocity dispersion with radius were only marginally
spectrally resolved (channel FWHM$_v$= 8.2 \kms), and likewise also
marginally spatially resolved ($5 < R_{HI}/{\rm FWHM}_{\theta} < 10$).

Of the \HI\ velocity dispersion measurements of high inclination galaxies
(see Table~\ref{tab:ch4-vel-disp-meas-highi}), only the study of
NGC4244 \citep{olling1996a} has the spatial resolution ($\sim$100 pc)
to investigate the radial velocity dispersion on scales relevant to
common galactic structure (star formation, spiral density waves, bars).
 However, Olling's method
(\citeyear{olling1996a}) used only part of the \HI\ emission
distribution, and may not be accurate (see
Sect.~\ref{sec:olling-RC-disp-meth}). The other high inclination
velocity dispersion measurements obtained by \citet{kvdkdb2004} used an
\HI\ velocity dispersion fit that was held fixed with radius, 
consequently no information was obtained about the radial variability 
of the velocity dispersion (see the discussion of the method in
Sect.~\ref{sec:fixed-disp-iter-envtrac-RC}).

As a result, very little is known about the velocity dispersion of the
\HI\ in the ISM. The dependence of the \HI\ velocity dispersion on
galaxy mass, surface brightness, star formation rate, and density
variations such as spiral arms or bars is unknown. Proper
modelling of the full \HI\ emission cube gives the radial distribution
of surface density, rotation and velocity dispersion. This approach
could be expanded to also model the \HI\ transparency of the \HI\
emission distribution of a galaxy, whether edge-on or inclined. 

In the past, low telescope resolution and deprojection issues made
such studies nonviable. In this paper, we show that it is possible to
accurately perform the deprojection required to measure the velocity
dispersion, rotation and surface density as functions of radius in
edge-on galaxies. The method is also applicable to galaxies of
moderate inclination. The edge-on orientation of galaxies in our study
is only required to minimise the uncertainty of the measured vertical
gas flaring, which is required to determine the dark halo flattening.
If fitting galaxies less inclined than edge-on, the inclination must
be accurately fitted at all radii, and the galaxy must not be too
close to face-on or else the line-of-sight velocity will be not be 
sufficiently resolved to model
the galaxy rotation. Application of our new \HI\ emission modelling 
method to nearby non-edge-on spiral galaxies, would allow investigation 
of the velocity dispersion, transparency,
temperature and density structure of neutral hydrogen in the ISM of a 
large sample of galaxies. 
This could explain how this small-scale ISM structure varies with the 
mass and star
formation rate of spiral galaxies.

Many methods have been devised to measure the rotation curves of
edge-on galaxies. In Table~\ref{tab:ch4-RC-meas}, we display the
previous rotation curve measurements of edge-on galaxies, showing the
method used and the resolution of the observations. In the next section, 
we will briefly review these methods, and the few previous
methods to measure the gas velocity dispersion and surface density,
before presenting our new method in Sect.~\ref{sec:iter-raddecomp-all3}.

In the paper III of this series 
we will present the results of applying this new
method to the \HI\ observations of the edge-on galaxy sample
presented in paper I. In that paper, we will paper also
derive the radial flaring of the \HI\ layer of these galaxies. 
A determination of the flattening
of the dark halo of one galaxy in our sample, UGC7321, will be 
presented in paper IV.

\section{Methods to extract rotation curves for edge-on galaxies}

High accuracy rotation curves and deprojected \HI\ surface densities
can be difficult to measure accurately from \HI\ observations of
edge-on disk galaxies, due to the projection of the velocity profiles
from a range of radii into a single line-of-sight velocity profile.
We will first discuss the methods that have been used previously and
the accuracy of their results. None of these methods independently {\it 
determine} the \HI\ gas velocity dispersion from the data. These methods 
provided a useful starting point from which to
develop a new method that incorporates the determination of velocity 
dispersion directly from the observations.

\subsection{The peak flux \&\ envelope tracing method for measuring
  the rotation curve}
\label{sec:peakflux-envtrac-RC}

Most previous measurements of the \HI\ rotation curves from edge-on
galaxies use the shape of
the outer envelope of the position velocity diagram, using either the
Peak Flux method \citep{rbft1985,mfb1992,um2003} or the
Envelope Tracing method \citep{sa1979,sofue1996,grsk2002}. The Peak
Flux method adopts the line-of-sight velocity at the maximum flux
along each slice of the XV diagram as the rotation velocity at the
line of nodes in that sightline. The Envelope Tracing
method instead uses the line-of-sight velocity at a nominal level,
e.g.  $20\%$ of the peak flux, on the outer velocity side of each
slice through the XV diagram.  Both these methods also involve application 
of a correction to the measured velocities, to correct for the 
combined line broadening due to the instrumental broadening and the 
unknown intrinsic velocity dispersion of the \HI.

The main problem with such methods is that the flux level of the
actual rotation velocity at any one radius is dependent on three
different unknown radial functions: the actual rotation curve, the
deprojected \HI\ surface density and the \HI\ velocity dispersion. The
peak flux method underestimates the rotation, while the envelope
tracing method tends to adopt the velocity at a flux level dependent
on the signal-to-noise of the observations. Too low a flux level will
overestimate the rotation, while the converse is true, if too high a
level is chosen.

The accuracy of derived rotation velocities also depends on the
accuracy of the adopted gaseous velocity dispersion used in the
velocity broadening correction of the measured velocity. Users
typically adopt a nominal radially invariant value of $6-10$ \kms\ from
direct vertical \HI\ velocity dispersion measurements of normal face-on
disk galaxies \citep{vdks1984,lsy1993}.

Beam-smearing also tends to increase the velocity of the outer
envelope, while lowering the velocity of the peak flux of slices in
the inner disk regions, where the radial gradient of rotation is larger.  
Beam smearing is a significant problem in most \HI\ rotation curve 
studies of edge-on spiral
galaxies as few measurements use observations with a resolution 
greater than $1-2$ kpc
(see Table~\ref{tab:ch4-RC-meas}).

The shape of the extreme velocity edge of the XV diagram is constructed
from superposition of gas all along the line-of-sight, including gas 
from the line-of-nodes to the edge of the disk. The gas at each location 
has an intrinsic
velocity dispersion that may be variable with radius. The edge-on 
orientation of the disk projects all of these radial
velocity profiles onto a single line-of-sight velocity profile at each
line-of-sight position through the disk \citep[see for example Fig.~1 of 
][]{kvdk2004}. 
Due to this projection effect it can be inferred that the envelope tracing 
method would fail to accurately measure the rotation curve, even if it 
were used with a correct \HI\ velocity dispersion, high-signal-to-noise 
and high resolution data.

Consequently, the rotation curve derived with these two methods
may exhibit an incorrect shape regardless of the signal-to-noise and
spatial resolution of the observations.  For example, an adopted \HI\
velocity dispersion say with an dispersion uncertainty of $3$ \kms,
will cause a rotation curve uncertainty of $7$ \kms\ due to the
velocity dispersion error alone. As the \HI\
velocity dispersion uncertainty is not known, if it is adopted from low
resolution face-on measurements of other spiral galaxies, the actual
rotation curve error could be larger. Consequently many high
inclination rotation curve measurement are likely to be in error by
between $7-20$ \kms, when the effects of projection and beam smearing
are included. Worse still, if the \HI\ were partially opaque, the inner
disk rotation may be further underestimated.

\begin{figure}[t]
\includegraphics[width=9cm]{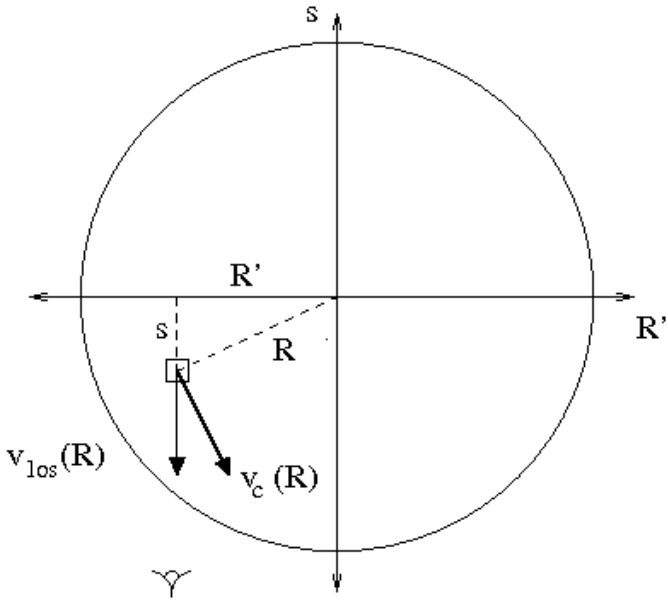}
  \caption[Diagram of galaxy coordinate system used to model the XV
    gas distribution]{Diagram of galaxy coordinate system used to model the XV
    gas distribution in Eqn.~(\ref{ch4-vel-profile}). This shows the
    major axis $R^{\prime}$, depth along the line-of-sight $s$, and
    the projection of galaxy rotation along the line-of-sight towards
    the observer.}
  \label{fig:ch4-coord_system}
\end{figure}

\subsection{The iterative envelope tracing method for measuring
  the rotation curve}
\label{sec:iter-env-trac-RC}

In 2002, \citet{ts2002} developed the iterative envelope tracing
method. This improves on the standard envelope tracing method by
automatically adjusting the initial envelope tracing rotation curve,
until the envelope tracing fit to a model XV distribution built with
this model rotation curve matches the original envelope tracing
rotation curve fit to the observed XV distribution. 
To test the rotation curve fit derived with the standard envelope tracing method, 
\citet{ts2002} used this fit to build a model XV distribution which they also 
fitted with the envelope tracing method. By comparing the two fits, and adjusting 
the model rotation curve accordingly, they incremented the model rotation curve 
until it generated an XV diagram that produced the same envelope tracing fit as 
the observed XV diagram. The initial fit $v_{fit,0}(R)$ to the observed XV 
diagram was used as the first model rotation curve $v_{model,1}(R)$, with 
which the first model XV diagram XV$_1$ was built and a new envelope tracing fit 
$v_{fit,1}(R)$ obtained. This was used to estimate a revised rotation curve
 model $v_{model,2}(R)$ by comparing the new fit $v_{fit,1}(R)$ to the original 
fit $v_{fit,0}(R)$. At each radius where the difference was greater than some 
tolerance $v_{crit}$, the model rotation curve was incremented by the difference 
$v_{fit,0}(R) -v_{fit,1}(R)$, thereby forming a revised model rotation curve 
$v_{model,2}(R)$. A new model XV diagram was built with the revised model rotation 
curve and the process was iterated, as shown below in Eqn.~(2)-(4), until the fit 
$v_{fit,n}(R)$ matched the original fit $v_{fit,0}(R)$. Whereupon the model 
rotation curve $v_{model,n}(R)$ was taken to be the best model for the observations.
\begin{eqnarray}
\label{eq:ch4-iter-env-trac.1}
\Delta v_{fit,n}(R) &=& v_{fit,0}(R) - v_{fit,n}(R) \\
\label{eq:ch4-iter-env-trac.2}
v_{model,n+1}(R) &=&  v_{fit,0}(R) + \Delta v_{fit,n}(R) \\
 & &  {\rm if }\  \Delta v_{fit,n}(R) >    v_{crit}, \nonumber \\ 
\label{eq:ch4-iter-env-trac.3}
v_{model,n+1}(R)  &=&  v_{fit,0}(R) \\
 & & {\rm if } \Delta v_{fit,n}(R) \leq  v_{crit}. \nonumber 
\end{eqnarray}

Takamiya \&\ Sofue employed several assumptions 
to build each model XV diagram. Firstly, the
radial gas surface density was approximated by the gas surface density along 
the major axis, which was derived from the observed XV diagram by integrating
over velocity. Secondly, the gas velocity dispersion was assumed to be
isotropic and isothermal in $R$ and $z$ with an adopted 
a value of $10$ \kms, which is typically observed in \HI\ in
spiral disks \citep{svdk1984, vdks1984}. In the
$z$ direction the \HI\ surface density was approximated by a 
$\textrm{sech}^2$ model 
with a constant scale height of $100$ pc. To simulate the observing 
process, each model 
XV cube was then convolved by the spatial and spectral resolution of the 
telescope, and 
integrated over $z$ to form the model XV map.

\citet{ts2002} applied this method and the earlier peak flux and
envelope tracing methods to optical emission spectra and CO (J=1-0)
observations with the Nobeyama Millimetre array on the edge-on galaxy
NGC3079, and four moderately inclined galaxies: NGC4536, NGC4254,
NGC4419, NGC4501.  They found that both the envelope tracing and peak
flux methods underestimated the rotation velocities in the inner
regions of their galaxies, with the peak flux method being in error by
up to $50-100\%$. Furthermore at outer radii with low signal-to-noise, the
derived rotation velocities exhibited dramatic scatter of $\pm 20-100$
\kms\ \citep[see Fig.~4 of ][]{ts2002}, probably because the $3\sigma$
noise level was above the $20\%$ flux level used for envelope tracing.

\citet{skno2003} applied the method to CO observations
of a sample of moderately inclined galaxies ($25^{\circ} < i < 74^{\circ}$), but
this time with a small difference.  Instead of fitting the XV diagrams at a
single fractional flux level, they fitted the line-of-sight velocity at a
series of flux levels $l$= 20, 40, 60, 80 and 100\%, and replaced
Eqn.~(\ref{eq:ch4-iter-env-trac.1}) by
\begin{equation}
  \Delta v_{fit,n}(R) = \sum_l g_l \times \left[ v^l_{fit,0}(R) -
   v^l_{fit,n}(R) \right]
\end{equation}
where $g_l$ is a set of weights satisfying $\sum_l g_l$= 1, and
usually evenly weighted, except when low flux levels are biased by
noise in sightlines through the XV diagram with low signal-to-noise.  The
iterative envelope tracing method is an improvement because it checks the 
veracity of the measured rotation curve. However, substantial flaws remain: 
it uses an incorrect radial gas surface density, and an assumed constant gas 
velocity dispersion.

\subsection{The fixed velocity dispersion, 1D Gaussian envelope
  fitting method for measuring the rotation curve \& approximating the
  gas velocity dispersion}
\label{sec:fixed-disp-iter-envtrac-RC}

\citet{grsk2002} improved on the standard Gaussian model of the full
outer envelope of the XV diagram, by performing a series of three-parameter
Gaussian fits to the outer envelope of the velocity profile in each
sightline. The first channel on the outer velocity envelope with flux $>
3\sigma$ was the outermost velocity channel used in each Gaussian fit. 
For the first Gaussian fit, this outer channel and the two
lower channels were included in the profile sample. Subsequent fits were
performed, each time adding the next lower velocity channel, until the
velocity channel with peak flux was reached. The centre of the Gaussian
with the smallest reduced $\chi^2$ was retained for that sightline.

\citet{kvdkdb2004} applied the fixed velocity dispersion 1D Gaussian
envelope fitting method to 14 galaxies using an radially constant
velocity dispersion of $10$ \kms. Successful fits were obtained for 11 
galaxies using this \HI\ velocity dispersion, however the remaining three 
galaxies required the velocity dispersion to be reduced to $6-7$ \kms\ to 
achieve a converging fit. However, the large beam size of the
observations ($1.7 \lesim {\rm FWHM}_{\theta} \lesim 18$ kpc) masked any small
scale variations in the rotation speed. Only 3 of the galaxies in
their sample have a \HI\ radius greater than $5$ beamwidths.
(see Table~\ref{tab:ch4-vel-disp-meas-highi}). For two of the three 
galaxies, which required smaller \HI\ velocity dispersions, the radius 
was spanned by only $2.5$ synthesized
beams. Consequently, due to the low spatial resolution, they were
unable to measure any radial structure of the velocity dispersion.

We show in Sect.~\ref{sec:ch4-disc-eff-beam} that even an intrinsically
constant gas velocity dispersion should appear broadened near the
Galaxy centre due to beam smearing --- typically by $\gesim 50-100\%$, if
FWHM$_{\theta} > 1$ kpc. This beam smearing effect would cause the
$\chi^2$ error of the velocity dispersion fits at inner radii to
dramatically increase as the XV slice bears progressively less
resemblance to a sum of Gaussians.

An advantage of this method is that it could be extended to measure
the gas velocity dispersion as a function of radius, by allowing the
velocity dispersion to be a free parameter in the Gaussian envelope
fits, and correcting the measured velocity dispersion for instrumental
broadening using Eqn.~(\ref{eq:ch4-instr-broadening-corr}).

\begin{figure*}[t]
\centering
\includegraphics[width=15cm]{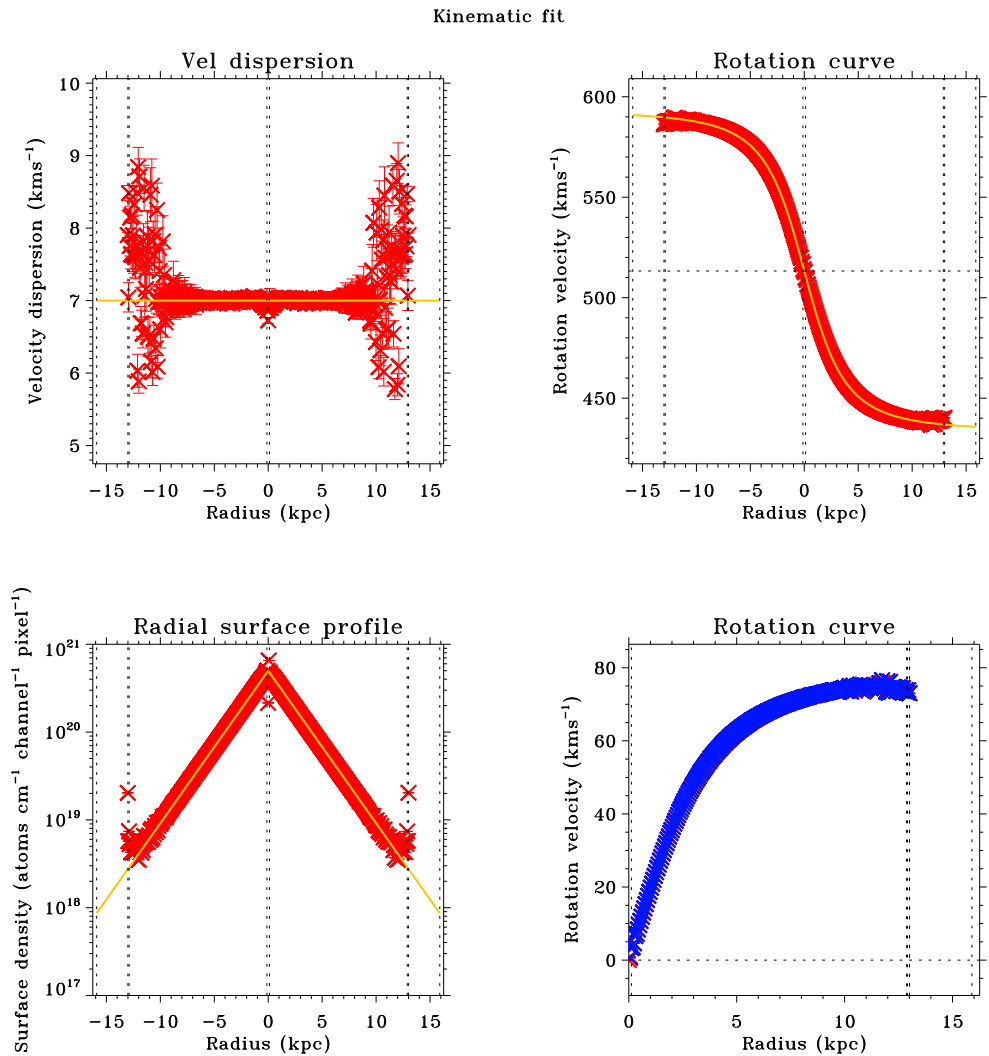}
  \caption[Gas kinematics fitted from a raw synthetic XV map using the
  radial decomposition method]{This is a derived gas kinematics fitted
    from a raw synthetic XV map (ie. not spatially smeared by the
    telescope beam and without noise) using the radial decomposition
    method. The plots of gas velocity dispersion, rotation and
    deprojected surface density show the actual synthetic galaxy
    kinematic as a grey (yellow) line and measured fits in (red)
crosses. The plot in the
    bottom right hand corner shows both sides of the measured rotation
    curve (in the color version the receding side is plotted first in red 
and the approaching side on top 
    in blue). Error bars are shown on all fits, though in most
    cases they are smaller than the plot symbol. Subsequent similar figures
in this paper will use the same plotting scheme.}
  \label{fig:ch4-raw-xvfit}
\end{figure*}

\begin{figure}[t]
  \centering
\includegraphics[width=9cm]{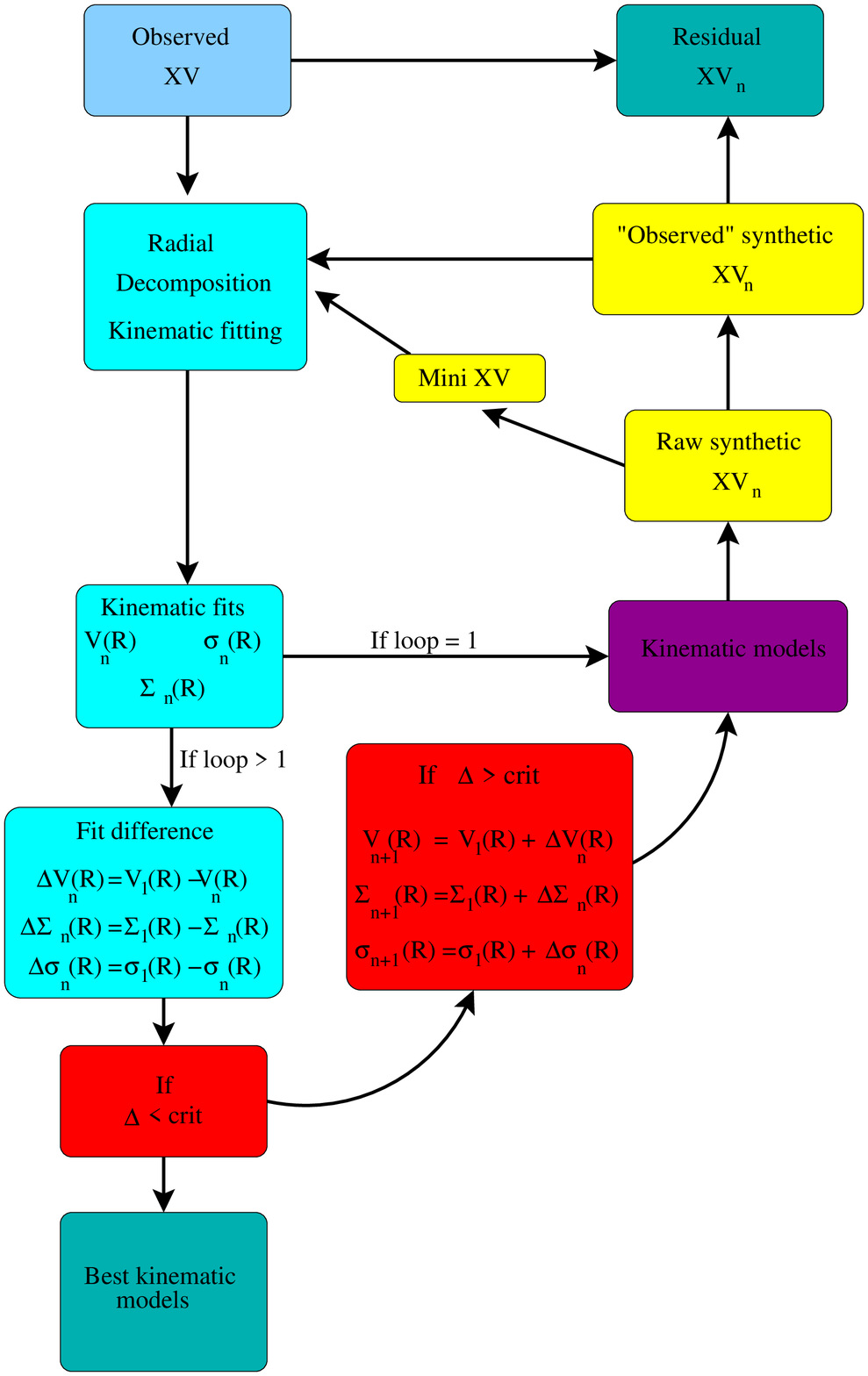}
  \caption[Flow chart showing the iterative radial decomposition
  method]{Flow chart showing the iterative radial decomposition method.}
  \label{fig:ch4-iter-flowchart}
\end{figure}

\subsection{The fixed velocity dispersion, 2D Gaussian envelope
  fitting method for measuring the rotation curve}
\label{sec:2D-fixed-disp-iter-envtrac-RC}

To attempt to correct the fixed velocity dispersion 1D Gaussian
envelope fitting method (Sect.~\ref{sec:fixed-disp-iter-envtrac-RC}),
\citet{grsk2002} applied the same series of fixed dispersion Gaussian
fits to slices along the radial $R^{\prime}$ direction of the XV diagram; 
i.e., in
addition to fitting at velocity profile at each major axis position
$R^{\prime}$, they also fitted major axis $R^{\prime}$ profiles of the
flux in each velocity channel $v$.  The width of the outer envelope in
the $R^{\prime}$ direction was determined by the FWHM$_{\theta}$ of the
synthesised beam and the unknown gradient of the rotation curve. They
measured this spatial dispersion along the $R^{\prime}$ profile at the
systemic velocity, $v_{los}$=$v_{sys}$, of each galaxy, and held this
value fixed when Gaussian fitting the $R^{\prime}$ profile of each
channel.  The rotation velocity at each radius was taken to be the minimum 
of the rotation velocities measured with each method. .

Using an adopted \HI\ velocity dispersion of $8$ \kms, \citet{grsk2002}
applied this method to measure the rotation curves of 24 edge-on ($i >
80^{\circ}$) galaxies (see Table~\ref{tab:ch4-RC-meas}). Beam-smearing
correction was important for this sample as most of their observations
had a FWHM beam of $1-2$ kpc, and most of the galaxies in their sample
were moderately resolved spatially with $5 \leq R_{\HI}/{\rm FWHM}_{\theta}
< 10$.

\subsection{The fixed velocity dispersion, XV modelling method for
  measuring the rotation curve and deprojected \HI\ surface density}
\label{sec:fixed-disp-raddecomp-RC-surf}

The basic approach of this method is to directly fit the full
observed distribution in the XV plane to derive the
rotation curve and the radial \HI\ distribution. It was first applied by 
\citet{vdkruit1981} and was further developed by \citet{kvdk2004}. 
We regard this method as the most powerful approach. 
\citet{kvdk2004} model the full observed XV
diagram as a set of concentric, coplanar rings, by progressively
measuring the rotation $v(R)$ from the outside in, or as they call it
the ``onion-peeling'' approach.

For a diagram of the coordinate system we refer to
Fig.~\ref{fig:ch4-coord_system}. By modelling the \HI\ layer as an
axisymmetric, transparent isothermal sheet in pure circular rotation,
the velocity profile probed by sight lines through the XV diagram is

\begin{eqnarray}
  \label{ch4-vel-profile}
f_{\HI}(R^\prime,v) &=&\frac{1}{\sigma_{v,tot} \, \sqrt{2\pi}}
\int ds \,\,\Sigma_{\HI}(R)  \\
& & {\rm exp}\biggl( -\frac{\left[v - v_c(R)\,\,(R^{\prime}/R)\right]^2}
  {2 \sigma_{v,tot}^2}\biggr), \nonumber 
\end{eqnarray}
where
\begin{equation}
v_{los}(R) = \left(\frac{R^{\prime}}{R}\right) \, v_c(R), \ \ 
R = \sqrt{R^{\prime 2} + s^2} \nonumber 
\end{equation}
and  $R^{\prime}$ is the projected radius on the sky, $s$ the depth
along the line of sight (i.e. $s$= 0  at the line-of-nodes), $R$ the
galactocentric radius, $\Sigma_{\HI}(R)$ the radial face-on surface
density profile and $\sigma_{v,tot}(R)$ the total velocity
dispersion; see Eqn.~(\ref{eq:ch4-tot-disp}). $v_c(R)$ is the rotation
velocity and $v_{los}(R)$ is the projection of the rotation velocity
onto the line-of-sight.  This method assumes a constant
\HI\ velocity dispersion $\sigma_{v,\HI}$ of $10$ \kms. The face-on
surface density $\Sigma_{\HI}(R)$ was derived independently using the
Warmels' (\citeyear{warmels1988}) method, which uses the
Lucy algorithm to invert an Abel integral
\citep[e.g.][Eqn.~(1B) - (59b)]{bt1987}.

At each sightline the \HI\ surface density and rotation on the line of 
nodes was fitted using a single Gaussian. By starting at the outermost 
projected radius 
$R^{\prime}$=$R^{\prime}_{outer}$ and using
at least 4 independent data points above a nominal noise flux level,
the rotation velocity was determined by a Gaussian 
fit to the outer envelope. The next inwards profile at
$R^{\prime}$=$(R^{\prime}_{outer} - \Delta R^{\prime})$ was then fitted
simultaneously for $(R^{\prime}_{outer} - \Delta R^{\prime})$ and
$R^{\prime}$, where $\Delta R^{\prime}$ denoted the annulus width,
which was set to
the pixel size of the observations. An initial estimate of the value of
the rotational velocity determined in the previous fit was used at
$R^{\prime}_{outer}$, while at the line-of-nodes $(R^{\prime}_{outer}
- \Delta R^{\prime})$ the 1D Gaussian envelope tracing value was
measured. Each line profile was fitted as a set of fixed dispersion
Gaussians with flux determined from the derived surface density,
yielding the rotation velocity at the line-of-nodes $v(R=R^{\prime})$
and new values for the rotation velocities (measured in projection) of
all high radii annuli in the sight line.

The procedure was repeated, progressively determining the rotation
velocity at all radii inwards until the centre was reached, and then
the same method was applied to the other side of the galaxy. In this
way the fitting algorithm does not become ill-constrained, as the
number of free parameters increases at the same rate as the initial
estimates of the fit parameters.

In addition to more accurately determining the rotation velocity by
modelling the full profile through each sight line of the XV diagram,
\citet{kvdk2004} noted that
this method also successfully corrects the rotation velocity
for the effects of beam smearing. This was done by first building a
model mini XV diagram comprising sufficient sightlines adjacent and
outwards of $R^{\prime}$ to span at least the beam FWHM$_{\theta}$ using
Eqn.~(\ref{ch4-vel-profile}), and convolving by the instrumental broadening
and any Hanning taper (if applied). The mini XV diagram is then convolved
spatially by the known beam FWHM$_{\theta}$. The velocity profile of interest,
can then be corrected for the modelled beam smearing effects
determined from the mini XV diagram, prior to Gaussian fitting.

The best fit rotation velocity at each radius was then determined by
$\chi^2$ minimization using the `amoeba' \citep{ptvf1992}
implementation of the simplex method \citep{nm1965}. This easily 
handles a large number of free parameters and does not require the 
derivatives of the model with respect to
those parameters (a computation-intensive calculation). Errors were estimated
using Monte Carlo simulations.

\citet{kvdk2004} applied this method to the \HI\ observations of 14
edge-on galaxies. These comprised 13 of the galaxies in the
\citet{kvdkdb2004} sample and \HI\ observations of NGC891 by
\citet{ssvdh1997}. The fitting method successfully converged to
sensible results for all the galaxies, except five galaxies with low
peak signal-to-noise of $\sn < 15$, and one galaxy NGC5529 which
displays a pronounced warp of the disk plane. The resolution of the
galaxies with successful fits are shown in
Table~\ref{tab:ch4-RC-meas}. Despite the large beam sizes ($1 \lesim
{\rm FWHM}_{\theta} \lesim 10$), two of these successfully fitted galaxies
(NGC891 and ESO435-G25) that were well sampled by the beam with
$R_{HI}/{\rm FWHM}_{\theta} > 10$, while four had moderate sampling with $5
\leq R_{\HI}/{\rm FWHM}_{\theta} < 10$, and two were successfully fitted
despite poor sampling of $R_{HI}/{\rm FWHM}_{\theta} < 5$.

For four galaxies with higher \sn\ observations, \citet{kvdk2004}
also succeeded in directly fitting the surface density by allowing it
to be a free parameter in the Gaussian fits. They compared
the rotation curve and surface densities derived from their method with
the rotation curve derived from standard envelope tracing method and
the surface density derived from the Warmels' (\citeyear{warmels1988})
deprojection method. Each method was applied to a simulated \HI\ XV
diagram of NGC2403, built with the ``true'' rotation curve and surface
density measured by \citet{fvmso2002} using extensive 3-D modelling,
showing that the fixed velocity dispersion XV modelling method
correctly measures both the rotation curve and the deprojected surface
density. Kregel \&\ van der Kruit's surface density deprojection 
method is a significant
improvement on that of  \citep{warmels1988}, which was
unable to recover the surface density variations on radial scales
smaller than the FWHM$_{\theta}$ of the synthesised beam.

\begin{subfigures}
\begin{figure*}[t]
  \centering
\includegraphics[width=9cm]{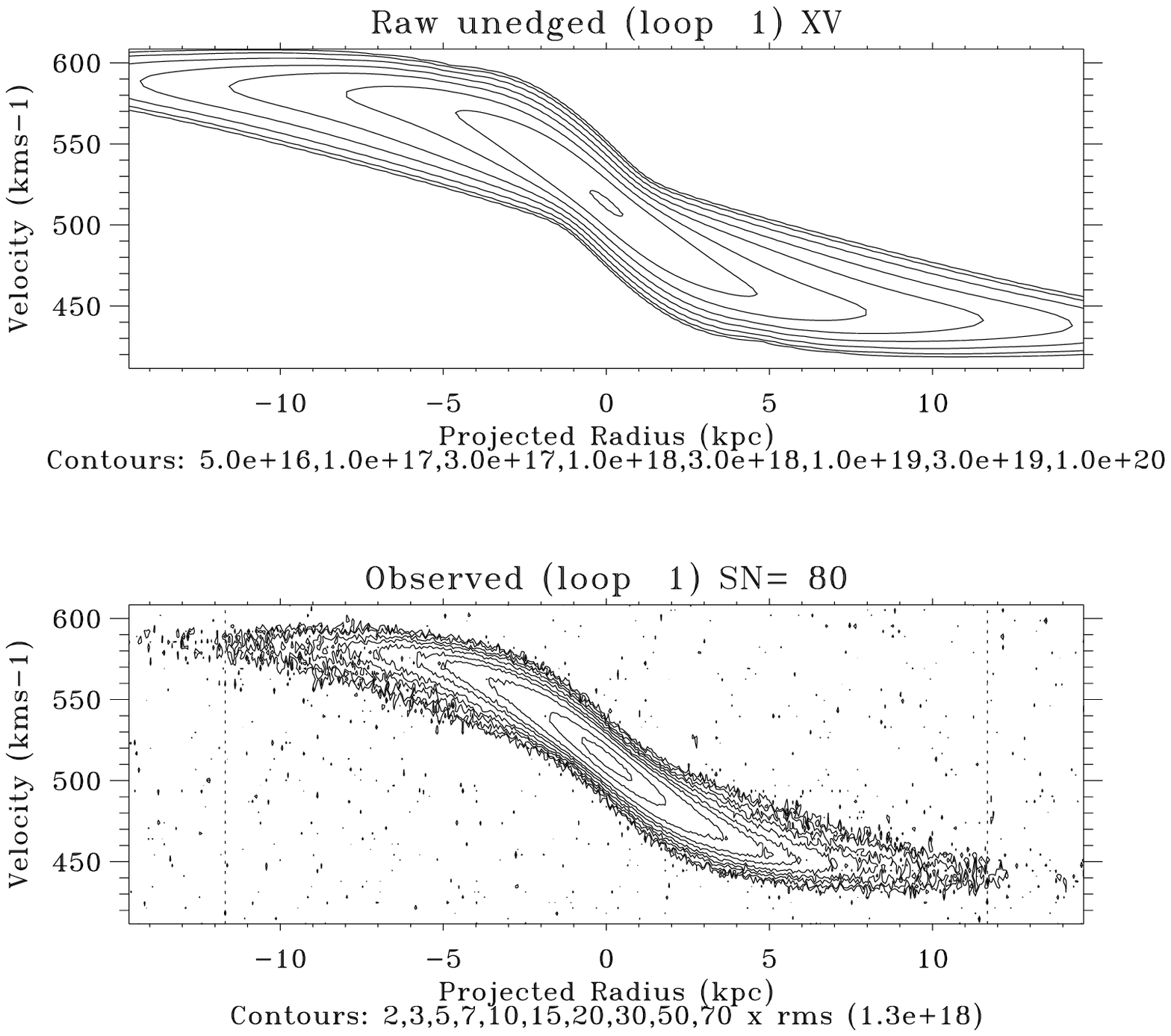}
\includegraphics[width=9cm]{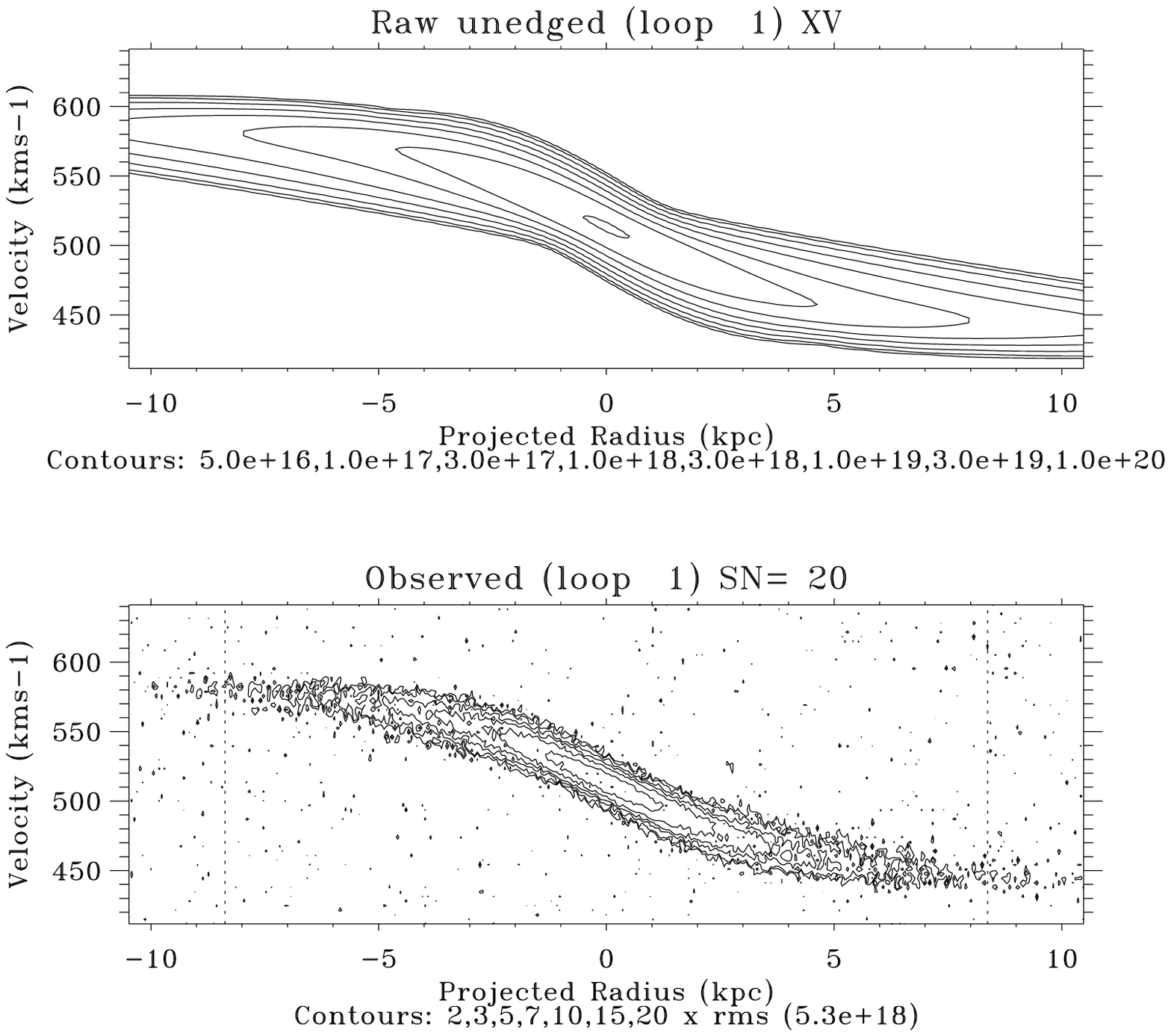}
  \caption[Effect of varying \HI\ sensitivity and synthesised \HI\ beam
  on the observed XV map]{Actual and observed XV maps of the synthetic
galaxy. This has the properties illustrated by the grey (yellow) lines in
Figs.~\ref{fig:smallbeam_kfit_SN80} and
\ref{fig:smallbeam_kfit_SN20}. The top XV diagrams show the
    actual XV \HI\ distribution, while the lower plot is the XV diagram as
    observed to detection limits corresponding to peak noise-to-noise
    ($S/N$) of 80 and 20, each with the same telescope FWHM of
    300 pc. The contours in the upper panels are (in units of
10$^{17}$ atoms cm$^{-1}$ pixel$^{-1}$)
at 0.5, 1, 3, 10, 30, 100, 300 and 1000 and in the lower panels at
2, 3, 5, 7, 10, 15,  20, 30, 50 and 70  times the r.m.s. noise
in the channel maps.
The effect of increased noise shows clearly the reduced
    fraction of the actual XV diagram that is
    detected. The vertical dashed lines on the observed XV maps mark
    the outermost radius with sufficient flux above $2\sigma$ to allow
    kinematic fitting. Note that the scales are different in
the plots on the right and the left, because of changing observable extent
for different S/N.}
  \label{fig:xv_raw_obs-diff_SN}
\end{figure*}

\setcounter{figure}{1}
\begin{figure*}[t]
  \centering
\includegraphics[width=15cm]{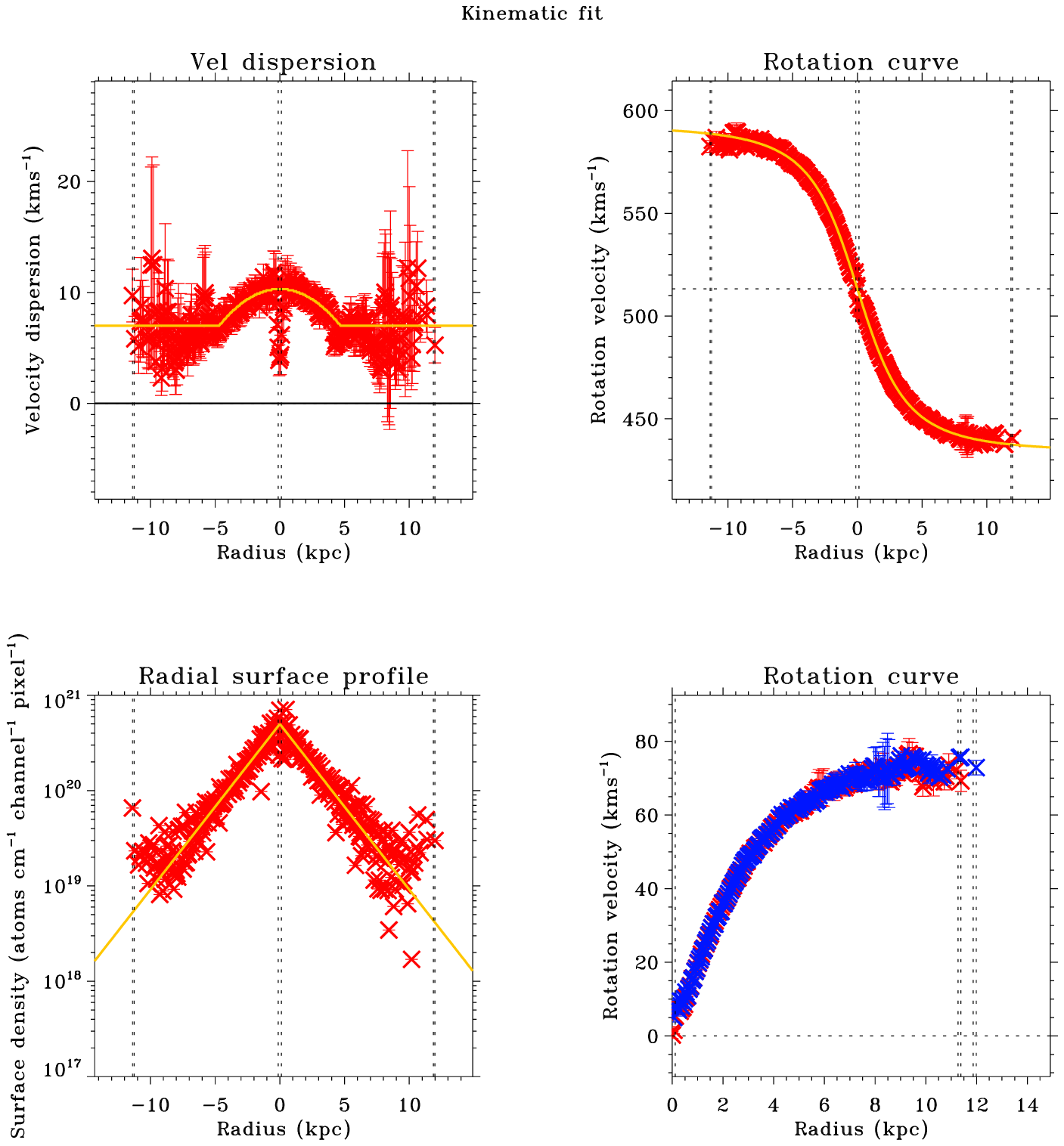}
  \caption[Measured kinematics of an XV diagram observed with a
  FWHM$_{\theta}$=300 pc telescope beam and a peak signal-to-noise of
  $80$.]{Fitted kinematics measured from an XV diagram observed with a
    telescope beam of FWHM$_{\theta}$=300 pc and to a peak
    signal-to-noise of $80$. Beam-smearing correction is not
    necessary at this high resolution, as the scatter due to
    beam-smearing is much smaller than the scatter due to noise. This
    figure and the plots in Fig.~\ref{fig:smallbeam_kfit_SN20} 
demonstrate the effect of noise on the
    measured kinematics. To reduce the scatter we have averaged the
    measurements using a evenly weighted 5 pixel wide bin after
    fitting the whole XV diagram via the radial decomposition method. The
    outermost dashed lines indicate the maximum radius $R_{max}$ with sufficient
    flux to be fitted. The other dashed vertical lines mark a beam
    HWHM from $R_{max}$ on either side, and a HWHM from the galaxy
    centre.}
  \label{fig:smallbeam_kfit_SN80}
\end{figure*}

\setcounter{figure}{2}
\begin{figure*}[t]
  \centering
\includegraphics[width=15cm]{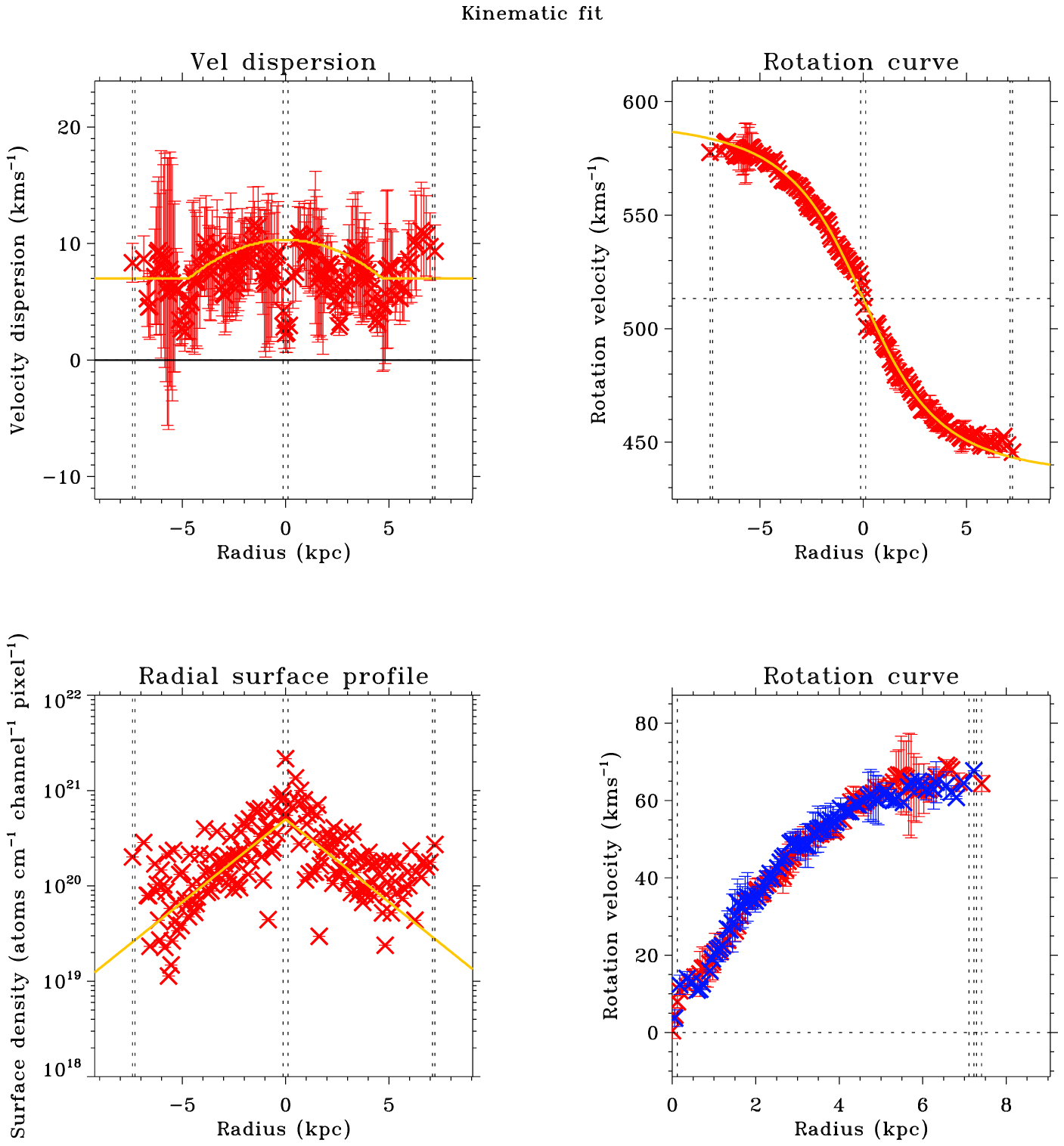}
  \caption[Measured kinematics of an XV diagram observed with a
  FWHM$_{\theta}$=300 pc telescope beam and a peak signal-to-noise of
  $20$.]{Fitted kinematics measured from an XV diagram observed with a
    telescope beam of FWHM$_{\theta}$=300 pc and to a peak
    signal-to-noise of $20$. Beam-smearing correction is not
    necessary at this high resolution, as the scatter due to
    beam-smearing is much smaller than the scatter due to noise. This
    figure and the plots in
    Fig.~\ref{fig:smallbeam_kfit_SN80} 
demonstrate the effect of noise on the
    measured kinematics. To reduce the scatter we have averaged the
    measurements using a evenly weighted 7 pixel wide bin after
    fitting the whole XV diagram via the radial decomposition method. The
    dashed line are as in Fig.~\ref{fig:smallbeam_kfit_SN80}.}
  \label{fig:smallbeam_kfit_SN20}
\end{figure*}
\end{subfigures}

\subsection{Direct outer envelope fit method for measuring the velocity
dispersion, rotation curve and deprojected \HI\ surface density}
\label{sec:olling-RC-disp-meth}

\citet{olling1996a} used the simple outer envelope velocity dispersion
fit, but then endeavoured to correct the measured broadening for the
major effects of beam-smearing and velocity profile projection in
addition to instrumental broadening. His correction method
\citep[detailed in Appendix A of ][]{olling1996a} calculated the
effect of the projection given in Eqn.~(\ref{ch4-vel-profile}) to
determine how to correct the measured direct Gaussian fit to the outer
velocity envelope of each slice of the XV diagram. By approximating the
Gaussian in Eqn.~(\ref{ch4-vel-profile}) as a parabola, and integrating
over several velocity channels around the line-of-node
($R$=$R^{\prime}$), Olling derived an analytical approximation of the
velocity profile around the line-of-nodes to be \citep[Eqn.~(A5) of
][]{olling1996a}:
\begin{eqnarray}
\label{eq:ch4-olling-approx}
  I(R^{\prime},v) &\approx & \frac{\epsilon_m K \Delta v
      \Sigma_{\HI}(R)}{\sigma_v(R) \sqrt{2 \pi}}  e^{-
    \frac{1}{2} \left(
      \frac{W(R)}{\sigma_v(R)} \right)^2} \times \nonumber \\
 & & \left( 2 - \frac{\epsilon_m^2 W(R) v_c(R)}{3
        \sigma_v(R)^2}
    \frac{R^{\prime} \epsilon_m^2}{3} \right. \nonumber \\
 & & \left. \left[ \frac{a
          W(R)}{\sigma_v(R)^2} 
        + b \left( \frac{W(R)^2}{\sigma_v(R)^2} - 1 \right) +
\right. \right. \nonumber \\
 & & \left. \left. 
      \frac{c}{\Sigma_{\HI}(R)} \right] \right),
\end{eqnarray}
where the kinematics (ie. $v_c(R)$, $\sigma_v(R)$ and
$\Sigma_{\HI}(R)$) were evaluated at the line-of-nodes $R$=$R^{\prime}$,
and $W(R)$=$v - v_c(R)$. $\Delta v$ is the channel width, $K$, $a$,
$b$ and $c$ are constants, $\epsilon_m$=$s_{max}/R^{\prime}$,
and $s$ is the depth along the line-of-sight (see the schematic of
the coordinate system in Fig.~\ref{fig:ch4-coord_system}). The first
term, $2$, results from gas on the line-of-nodes. The second term arises
due to material close to the major axis, and the last 3 terms arise
due to radial gradients in the rotation velocity, gas velocity
dispersion and surface density, respectively.

\citet{olling1996a} applied this method to one galaxy, NGC4244, which
was observed at high spatial and spectral resolution with a
FWHM$_{\theta}$ of $170$ pc and a FWHM channel width of $5.2$ \kms.
Although NGC4244 is a small Scd galaxy with maximum \HI\ radius of $11$
kpc, the high spatial resolution allowed the radial \HI\ structure to
be sampled by $63$ FWHM beams on either side of the galaxy centre (see
Table~\ref{tab:ch4-vel-disp-meas-highi}).  By testing the above method
on XV simulations of NGC4244, he found that
Eqn.~(\ref{eq:ch4-olling-approx}) was sufficient to recover the
kinematics in the outer galaxy. However, within $6$ kpc from the
galaxy centre, beam-smearing became a problem. To derive the gas 
properties in the inner disk consistent with the beam smearing, 
\citet{olling1996a} used simulations of NGC4244 to determine an 
empirical correction to the
kinematics measured with Eqn.~(\ref{eq:ch4-olling-approx}). In order to
simulate the XV diagram of NGC4244, Olling had to estimate the intrinsic
kinematics, as the derived corrections depend on the intrinsic XV diagram as
well as the resolution of the observations. With this method Olling
found that the velocity dispersion of NGC4244 was roughly constant at
$8.5\pm2$ \kms. Unlike the \HI\ velocity dispersion measurements of
face-on galaxies which mostly decline with radius (see
Table~\ref{tab:ch4-vel-disp-meas-lowi}), Olling's analysis of NGC4244
shows no radial structure in the velocity dispersion despite being
measured on scales of $\sim$100 pc.

\begin{subfigures}
\begin{figure*}[t]
 \centering
\includegraphics[width=15cm]{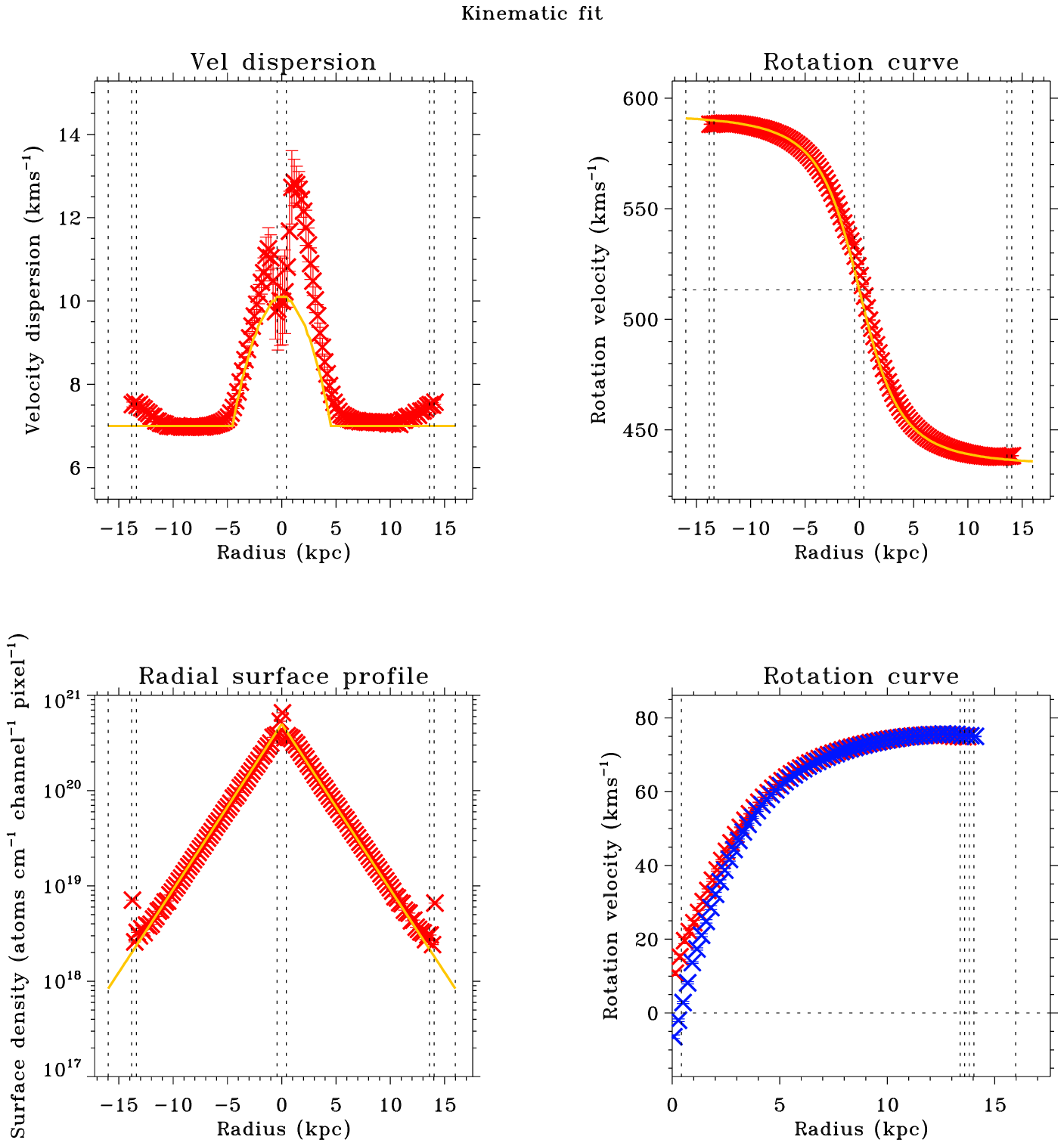}
  \caption[Fitted kinematics of a noiseless XV diagram observed with a
  FWHM$_{\theta}$=1.0 kpc telescope beam]{Fitted
    kinematics (without beam correction) measured from a noiseless XV diagram
    observed with a telescope beam of FWHM$_{\theta}$=1 kpc.  The
    plots shows that the rotation velocity exhibits a small
    beam-smearing effect and the velocity dispersion is broadened due
    to beam-smearing by $\sim$50\% at the Galaxy centre. The
    dashed line are as in Fig.~\ref{fig:smallbeam_kfit_SN80}.}
  \label{fig:beameff_kfit_1.0beam}
\end{figure*}

\setcounter{figure}{1}
\begin{figure*}[t]
  \centering
\includegraphics[width=15cm]{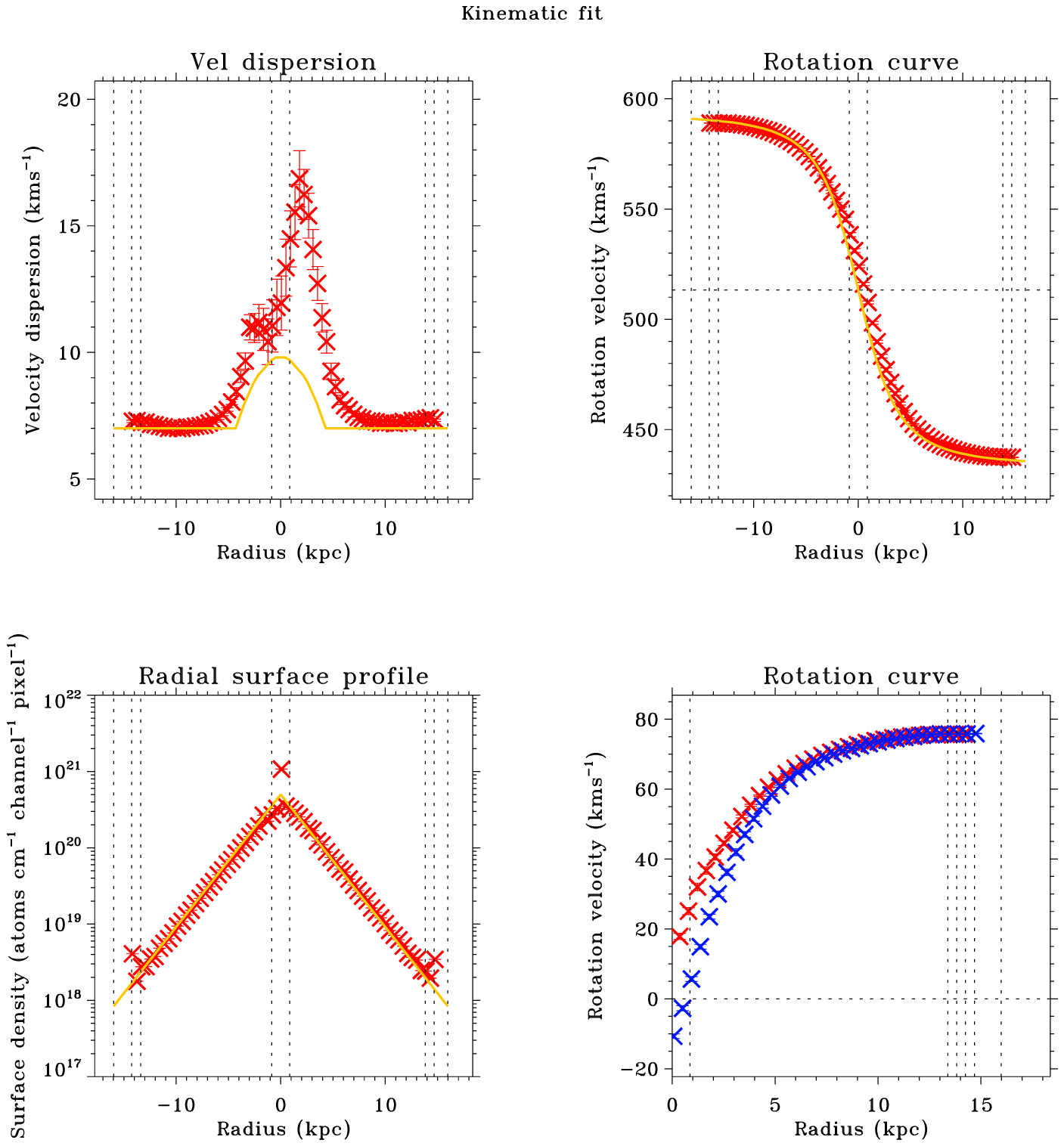}
  \caption[Fitted kinematics of a noiseless XV diagram observed with a
  FWHM$_{\theta}$=2 kpc telescope beam]{Fitted kinematics
    (without beam correction) measured from a noiseless XV diagram observed
    with a telescope beam of FWHM$_{\theta}$=2.0 kpc.  The plots shows
    that beam-smearing by a $2$ kpc beam causes a large error in the
    inner rotation velocity and a $\sim$70\% increase of the measured
    central velocity dispersion. The
    dashed line are as in Fig.~\ref{fig:smallbeam_kfit_SN80}.}
  \label{fig:beameff_kfit_2.0beam}
\end{figure*}
\end{subfigures}

\section{The radial decomposition XV modelling method}
\label{sec:iter-raddecomp-all3}

We developed the radial decomposition XV modelling method to more
accurately measure the rotation, velocity dispersion and deprojected
surface density of \HI\ disks.  Like the \citet{kvdk2004} method
it operates by fitting the galaxy from the outer radii
inwards - ``onion-peeling'' as they called it. The radial
decomposition method works by progressively fitting each sightline of
the observed XV diagram, starting at the outermost edge, to measure the
rotation velocity, surface density and velocity dispersion. The basic
method is essentially the same as in \citet{kvdk2004}, however we also fit the
\HI\ velocity dispersion, and extend the method to iterate further if
required.  By progressing inwards, the method uses the measured
kinematics of outer radii to isolate and fit only the velocity profile
of the line of nodes gas along each sightline.

This allows one to accurately fit the rotation, velocity dispersion
and surface density as functions of galactocentric radius despite the
edge-on orientation of the galaxy. The measured velocity dispersion 
is corrected for the broadening
due to the finite spectral resolution and the derived \HI\ velocity
dispersion at each radius is used in the measurement of all inner
annuli.

The method works by modelling the radial and velocity distribution of
each slice through an edge-on galaxy that has been integrated over the
vertical axis $z$. This requires the usual assumptions of circular 
motion, azimuthal symmetry, 
isotropy of the \HI\ velocity dispersion tensor. To maximise the peak 
signal-to-noise of the XV
map, the map is formed by integrating over the full range $z$ spanned
by the edge-on galaxy in the vertical direction. This requires that
the velocity dispersion is vertically isothermal and that the rotation is
cylindrical.  In future studies, the assumption of \HI\ isothermality 
should be tested by applying the radial decomposition XV modelling method 
to deep observations of both low and high lattitude \HI\ observed at 
high spatial resolution. This would determine whether  the velocity
dispersion (and rotation) varies with height above the galactic plane.
Although the \HI\ in some deeply observed galaxies like NGC 891 
\citep{foss05} does show a rotational lag in the \HI\ layer
a few kpc above the galactic plane, our data is for the most part
confined to a thin layer of FWHM $<$ 2 kpc, so the weighted effect of the
high-latitude lag on our analysis is unlikely to be significant.

Although the method will not detect spatial asymmetries in the gas
surface density, such as might be expected from bars or spiral waves,
it will detect radial annuli devoid of \HI, or a central \HI\ hole,
as the surface density is modelled as $\Sigma_{\HI}(R)$.

The method fits the radial kinematics by modelling each sightline
through the XV diagram as a cut through a superposition of annular rings, in
which the gas at each radius exhibits a Gaussian velocity profile
described by the radial gas kinematics seen in projection according
to Eqn.~(\ref{ch4-vel-profile}). 
Each slice comprises gas at a range of radial bins, with the gas in
each radial bin in the slice forming a Gaussian profile\\
\indent
$\bullet $ centred at the rotation velocity seen in projection, $v_{los}$
($v_{los}$=$v(R)$ along the line-of-nodes, $v_{los} < v(R)$ for radii
outside the nodal radius);\\
\indent
$\bullet $ broadened to a velocity FWHM due to the effect 
  of intrinsic velocity dispersion and the instrumental resolution, 
  FWHM$_{v,tot}$ and\\
\indent
$\bullet $ with the integrated flux of each Gaussian consisting of the
  flux at that radius.

Starting at the outermost radius on one side of the galactic centre,
each line-of-sight is modelled as a sum of Gaussians, where each Gaussian 
simulates the gas at that radius. Due to the approximately
exponential outer \HI\ distribution of most galaxies, the gas from
radii near the line of nodes dominates the flux in each velocity
profile slice through the XV diagram. At the outermost line-of-sight, the
velocity profile is approximated as a single Gaussian as the flux can
be approximated as coming from a single radial bin with
$R^\prime$=$R_{max}$. The properties $v(R)$, $\sigma_{v,\HI}(R)$,
$\Sigma_{\HI}(R)$ at that radius can be directly measured at
$R_{max}$. In the next innermost radius, $R^\prime_2$=$R_{max}-(1
\textrm{pixel})$, the observed velocity profile is modelled as a
superposition of the Gaussian with the known $R_{max}$ kinematics, and
the Gaussian with kinematics from the line-of-nodes at radius
$R^\prime_2$ which is fitted for rotation, velocity dispersion and
total flux. By proceeding inwards in this manner, the model never
becomes ill-constrained as there are only 3 free parameters in the
fit to any sightline.

The radial kinematics are thus obtained with radial sampling
corresponding to the pixel size of the \HI\ image cube along the galaxy
major axis $R^{\prime}$.  
The derived kinematics from fitting to all slices of the raw XV diagram using
in the above slice fit are shown in Fig.~\ref{fig:ch4-raw-xvfit}.
The fit has been made to a synthetic galaxy with a constant
velocity dispersion, a simple rotation curve and an exponential surface
density distribution. 
The plot demonstrates that the radial decomposition method does an
excellent job of recovering the actual galaxy 
structure and kinematics. 
Near the edge of the edge-on galaxy, the fits are
systematically high in both the velocity dispersion and the surface
density, because the model assumes that the outermost detected slice
contains flux at a single radial bin. The overestimation of the
velocity dispersion, causes the fitting process to compensate on the
next few inner radii slice fits; thus the velocity dispersion fit
oscillates around the actual value in the outer few slices. The small
deviation in the centre slice is a numerical artifact. We remove this
artifact by excluding the central fits within $0.5$ kpc of the galaxy
centre. Although these plots show the derived fits to the XV diagram formed
from a simple galaxy model with a constant velocity dispersion, simple
surface density and rotation curve, the same fitting success is
achieved for fits to synthetic galaxies with more complicated
intrinsic kinematics.

\subsection{Instrumental broadening correction}
\label{sec:instru-broad-corr}

Fits of real galaxy observations also require correction for the 
instrumental broadening due to the signal response
of the spectral channels of the interferometer correlator.  For
simplicity, the signal response function of the correlator channels is
approximated as a Gaussian. Consequently, the dispersion due to both 
the \HI\ velocity dispersion and the instrumental broadening is the 
quadratic sum
of each property. Thus, the measured velocity
dispersion is related to the intrinsic \HI\ velocity dispersion and the
dispersion $\sigma_{v,instr}$ of a spectral channel with 
FWHM$_{v,instr}$=$2 \sigma_{v,instr} \sqrt{2 \log 2 }$ by:
\begin{equation}
  \label{eq:ch4-tot-disp}
  \sigma_{v,tot} = \sqrt{ \sigma_{v,\HI}^2 + \sigma_{v,instr}^2 }
\label{eq:ch4-instr-broadening-corr}
\end{equation}
This allows the intrinsic \HI\ velocity dispersion to be easily
derived from each Gaussian fit.

\begin{subfigures}
%
%
\begin{figure*}[t]
  \centering
\includegraphics[width=15cm]{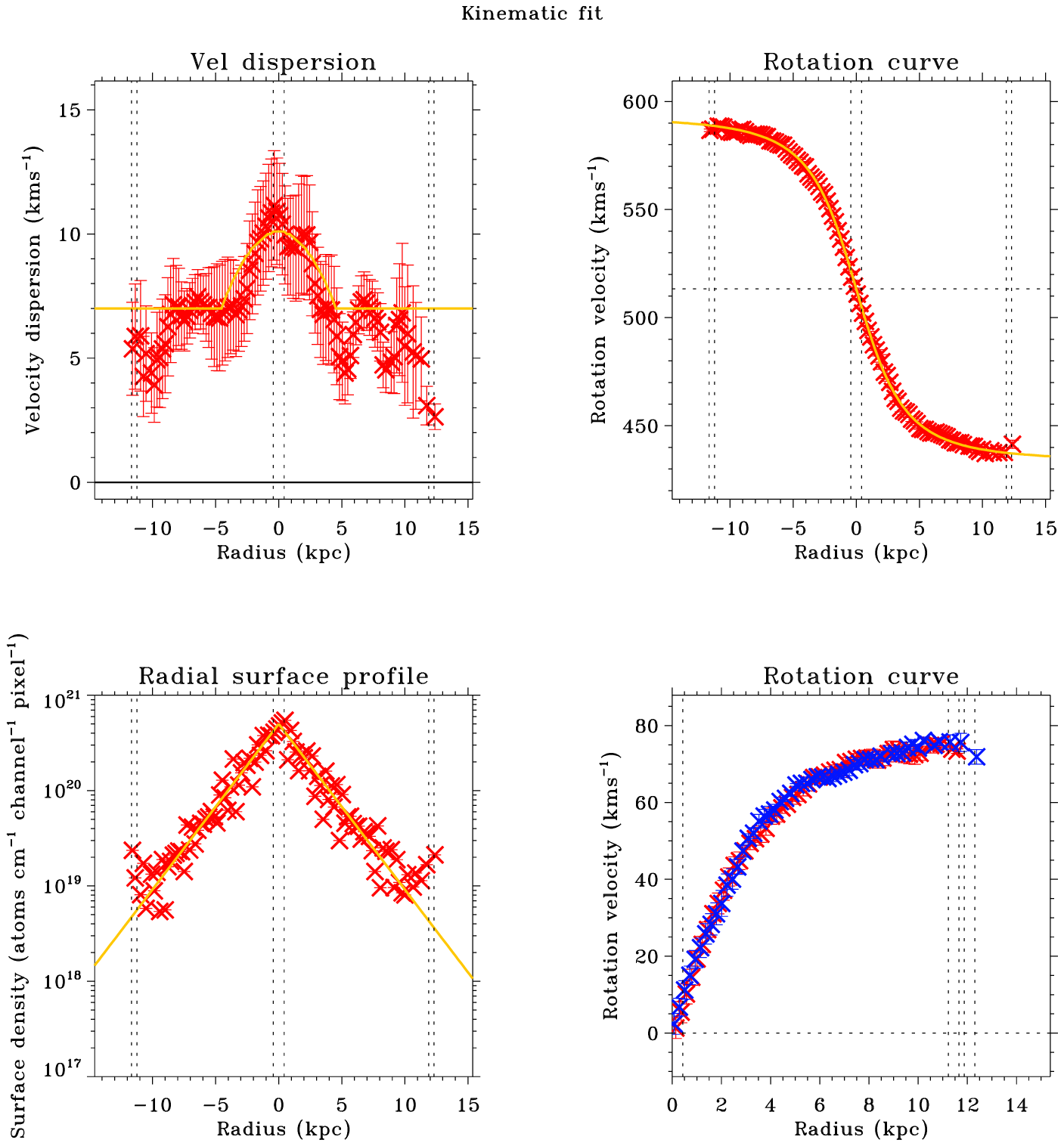}
  \caption[Measured kinematics of an XV diagram observed with a
  FWHM$_{\theta}$=1 kpc telescope beam and a sensitivity
  $S/N_{max}$ of 80]{Fitted kinematics (with beam correction)
    measured from XV diagram observed with a telescope beam of
    FWHM$_{\theta}$=1 kpc and with a peak signal-to-noise of 80. To
    reduce the scatter in the measured kinematics, the velocity
    dispersion and rotation curve were averaged (post fitting) over an
    evenly weighted bin of width of 5 pixels. In
    Fig.~\ref{fig:beamcorr_xv_1.0beam_SN80}, the XV diagram built with the
    best-fit measured kinematics and observed to the same telescope
    specifications is compared to the original XV diagram.}
  \label{fig:beamcorr_kfit_1.0beam_SN80}
\end{figure*}

\setcounter{figure}{1}
\begin{figure*}[h\t]
  \centering
\includegraphics[width=15cm]{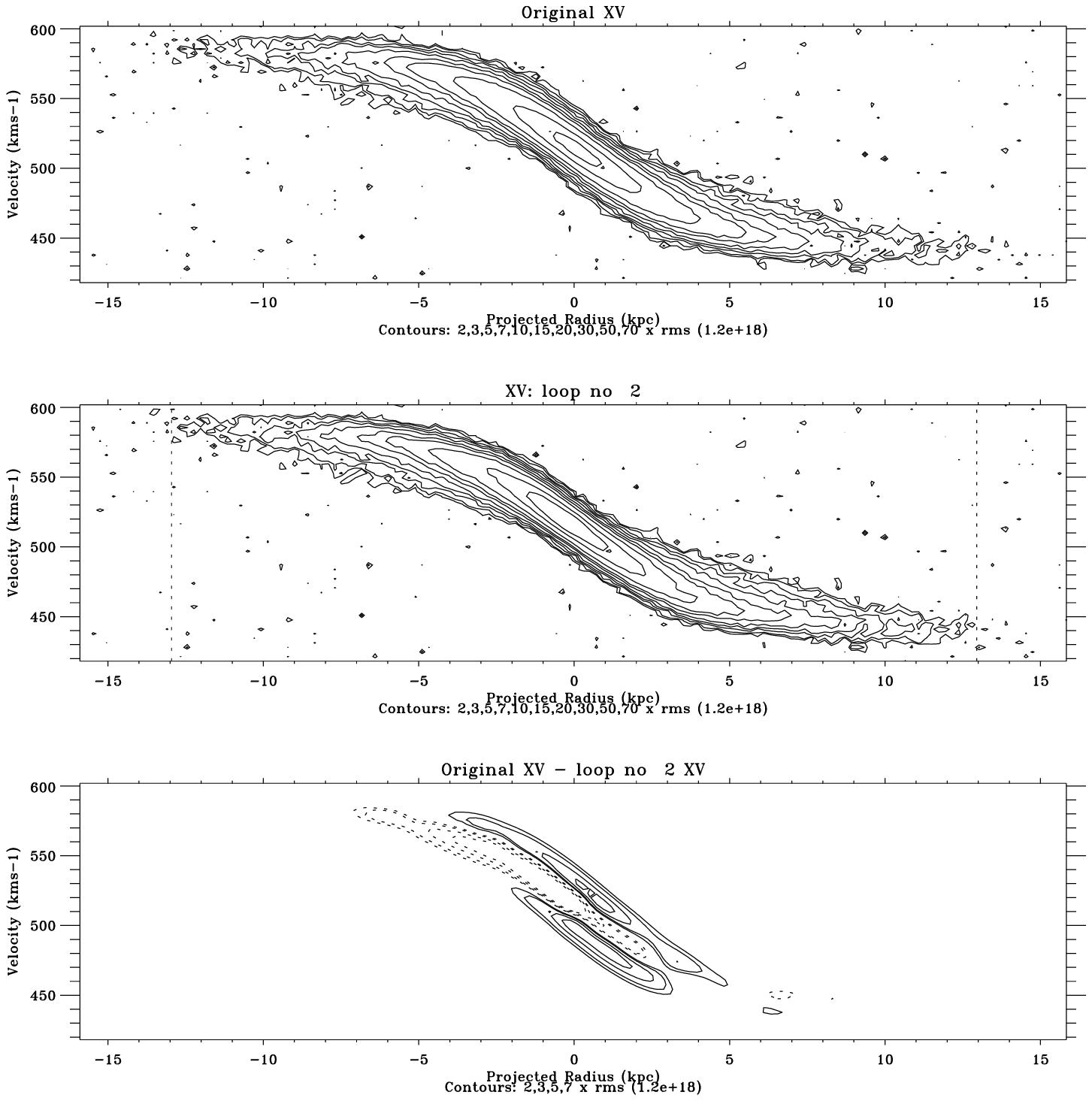}
  \caption[XV diagram (observed with a FWHM$_{\theta}$=1 kpc telescope beam
  and peak $S/N$ of 80) and XV diagram built with the measured kinematics]{Top:
    Original XV diagram observed with a FWHM$_{\theta}$=1 kpc telescope beam
    and peak $S/N$ of 80. Middle: The XV diagram built using the best-fit
    measured kinematics and observed with the same telescope
    resolution and sensitivity limits. The measured kinematics used to
    build the middle plot are shown in
    Fig.~\ref{fig:beamcorr_kfit_1.0beam_SN80}.
Bottom: The difference between the top and middle diagrams. Contour units are
2, 3, 5, 7, 10, 15, 20, 30, 50, 70 times the r.m.s. noise in the
channel maps.}
  \label{fig:beamcorr_xv_1.0beam_SN80}
\end{figure*}
\end{subfigures}

\begin{subfigures}
\begin{figure*}[t]
  \centering
\includegraphics[width=15cm]{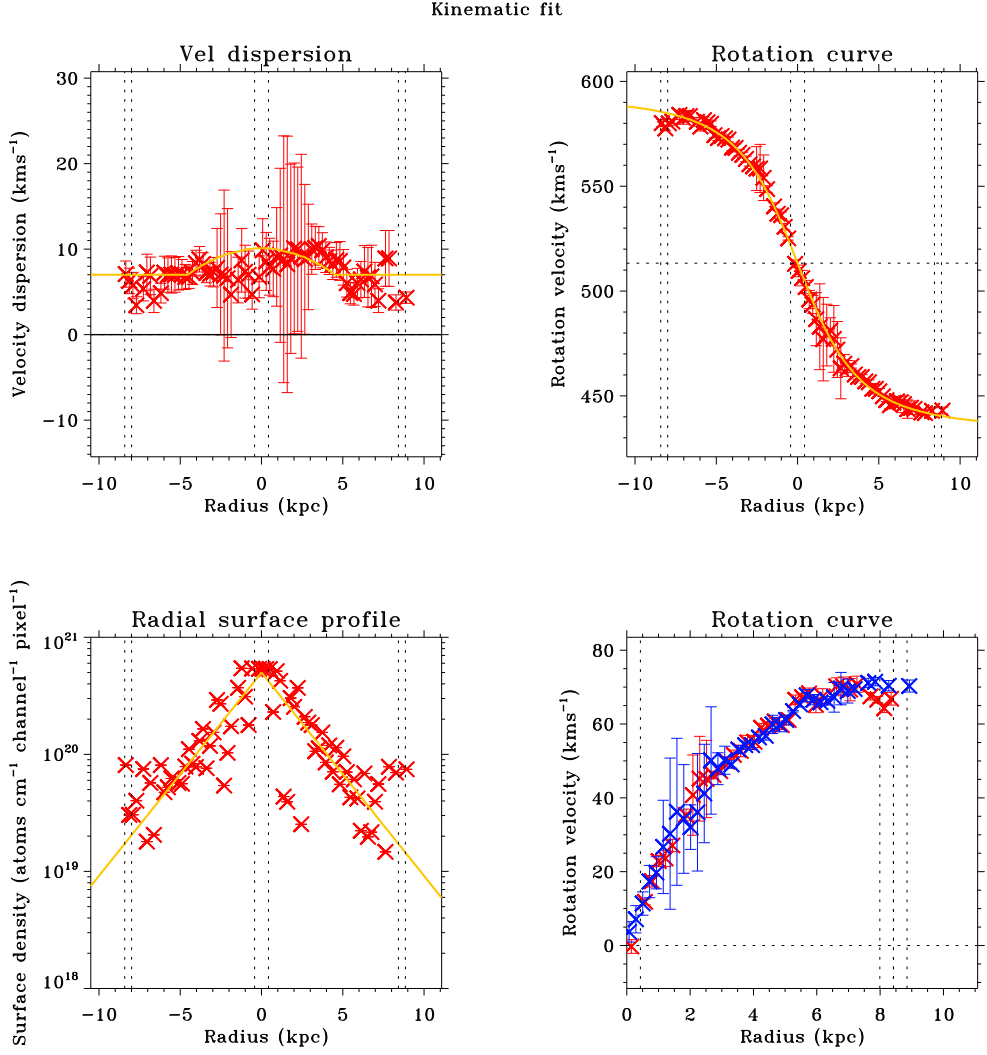}
  \caption[Measured kinematics of an XV diagram observed with a
  FWHM$_{\theta}$= 1 kpc telescope beam and to a sensitivity
  $S/N_{max}$ of 30]{Fitted kinematics (with beam correction)
    measured from XV diagram observed with a telescope beam of
    FWHM$_{\theta}$= 1 kpc and with a peak signal-to-noise of 30.
    Although the rotation velocity and the surface density are well
    recovered, the plot of the measured \HI\ velocity dispersion shows
    that a peak signal-to-noise of 30 is not high enough to perform
    beam-smearing correction for this kinematic model and moderate
    spatial resolution. The measurements are binned over 
    3 pixels post-fitting, however averaging only improves the
    measurement when it is accurate. By comparison, the observation
    with a larger beam to the same signal-to-noise recovers more of
    the \HI\ emission distribution and thus recovers the radial shape
    of the \HI\ velocity dispersion (compare to
    Fig.~\ref{fig:beamcorr_kfit_2.0beam_SN30}). In
    Fig.~\ref{fig:beamcorr_xv_1.0beam_SN30}, the XV diagram built with the
    best-fit measured kinematics and observed to the same telescope
    specifications is compared to the original XV diagram.}
  \label{fig:beamcorr_kfit_1.0beam_SN30}
\end{figure*}

\setcounter{figure}{1}
\begin{figure*}[t]
  \centering
\includegraphics[width=15cm]{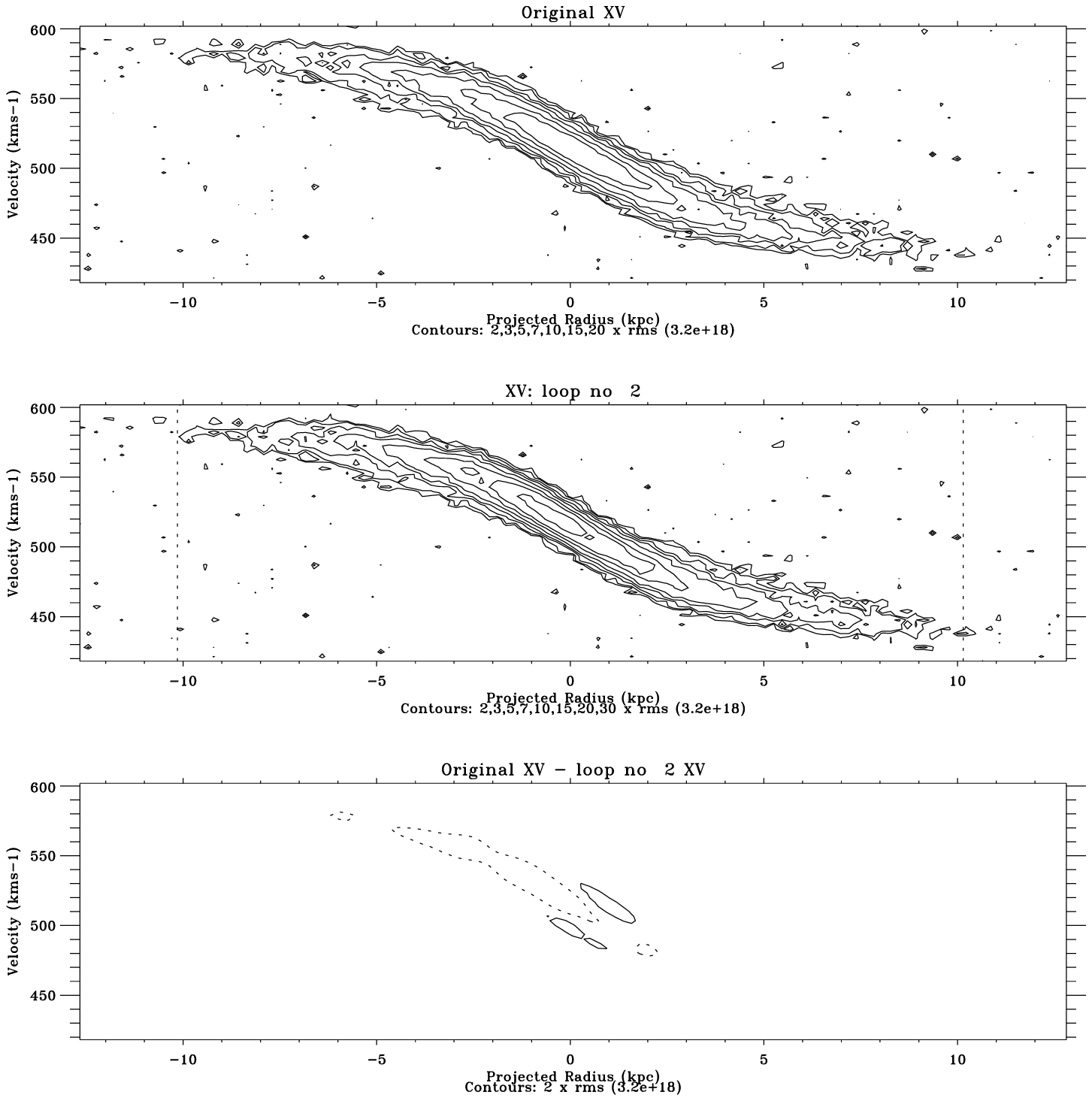}
  \caption[XV diagram (observed with a FWHM$_{\theta}$= 1 kpc telescope beam
  and peak $S/N$ of 30) and XV diagram built with the measured kinematics.]{Top:
    Original XV diagram observed with a FWHM$_{\theta}$= 1 kpc telescope beam
    and peak $S/N$ of 30. Middle: The XV diagram built using the best-fit
    measured kinematics and observed with the same telescope
    resolution and sensitivity limits. The measured kinematics used to
    build the middle plot are shown in
    Fig.~\ref{fig:beamcorr_kfit_1.0beam_SN30}. Bottom: 
The difference between the top and middle diagrams. Contour units are
2, 3, 5, 7, 10, 15, 20, 30 times the r.m.s. noise in the
channel maps.}
  \label{fig:beamcorr_xv_1.0beam_SN30}
\end{figure*}
\end{subfigures}

\subsection{Beam correction}
\label{sec:beam-corr}

Any method that aims to recover spatial resolution on scales of the
image pixels must also correct for the convolution effect of the 
telescope synthesised beam. The beam is approximated an
elliptical Gaussian, which is near circular for objects at an appropriate 
declination relative to the
latitude of the observatory. The telescope
beam causes a smearing effect along the spatial axes of the \HI\
channel map cube by convolving the flux distribution with a 2D
Gaussian. As the XV map is formed by integrating over $z$, then
providing the beam is near circular, the beam effectively smears the
spatial axis of the XV map by a 1D Gaussian with FWHM equal to that
of the beam, along the $R^{\prime}$ axis.

Beam smearing is a problem when the galaxy kinematics vary strongly on
intervals along the major axis smaller than the FWHM of the
synthesised beam. In practise, the largest variations usually 
occur in the inner region where the rotation curve rises steeply
and to a lesser extent in any region with 
variable kinematics due to heating processes in the ISM.

In the following section (\ref{sec:iter-process}), we describe
in detail how we incorporate an iterative process to check the derived
kinematics measured from the observed XV map. Briefly the process works
as follows. Firstly, the first-pass kinematics from the  the kinematic
model is used to build a synthetic galaxy XV diagram using
Eqn.~(\ref{ch4-vel-profile}), which is then convolved by
the known telescope resolution to simulate the actual observational
effects. This simulated XV diagram is then fitted with the radial decomposition
method and the resulting kinematics compared to the original fitted
kinematics. Where the difference between the kinematic fits is greater
than a critical level the kinematic models are incremented and the
process repeated until the fitted kinematics and the simulated galaxy
resemble the original fitted kinematics and the observed galaxy
distribution.

Initially beam correction was attempted by iterating the fitting
process as described above. However the velocity dispersion fits are
particularly sensitive to noise, and it was found to be difficult to
correct the velocity dispersion by iterative fitting if the first pass
fits were too deviant from the true fit, as often occurs when fitting
galaxies with steep inner rotation curves that have been observed with
a large beam.  The optimal method was found to be using the beam
correction method applied by \citet{kvdk2004} whereby instead of
fitting the actual observed slice, the slice is first corrected by
subtracting the flux contamination due to beam-smearing of flux in
nearby slices. The residual profile is then fitted using the radial
decomposition method. As the radial decomposition method fits
progressively inwards, the kinematics are known at radii larger than
the slice of interest $R^{\prime}$, thus it is possible to form a raw
``mini-XV diagram'' comprising solely the slices at higher $R^{\prime}$.
However to correctly derive the flux contamination due to beam
smearing a raw ``mini-XV diagram'' must be built spanning sufficient slices on
both sides, but excluding the slice of interest. This mini-XV diagram is
convolved along $R^{\prime}$ by a Gaussian of FWHM equal to that
of the beam FWHM$_{\theta}$ along $R^{\prime}$, to determine the
beam-smearing component of the flux in the slice of interest.

Consider the observed beam-smeared velocity profile at the slice of
interest $F_0$. This velocity profile is the sum of the 
intrinsic profile $f_0$
weighted by the beam centre $b_0$ and the flux of all nearby
slices weighted by their respective beam weights $b_n$, where $b_n$
becomes negligible for slices further than a few FWHM$_{\theta}$'s
from the slice of interest. Thus
\begin{equation}
  \label{ch4-beamconv-slice}
  F_0 = f_0 b_0 + \sum_{n=1}^{\infty} f_n b_n + \sum_{n=-1}^{-\infty}
  f_n b_n.
\end{equation}
Here each beam weight $b_n$ is the flux of a normalized Gaussian in
the $n^{th}$ pixel from the beam centre. Therefore, the intrinsic
velocity profile is
\begin{equation}
  \label{ch4-beamcorr-slice}
  f_0 = \frac{ F_0 - \left( \sum_{n=1}^{\infty} f_n b_n +
      \sum_{n=-1}^{-\infty} f_n b_n \right) }{b_0},
\end{equation}
where $b_0$ is the peak flux of a normalised Gaussian beam
\begin{equation}
  b_0 = \frac{1}{ \sqrt{2 \pi} \sigma_{\theta} } \ \ {\rm with}\ \ 
\sigma_{\theta} = \frac{{\rm FWHM}_{\theta}}{ 2 \sqrt{2\log{2}} }.
\end{equation}

Beam correction is thus undertaken by determining $\sum_{n=1}^{\infty}
f_n b_n$ and $\sum_{n=-1}^{-\infty} f_n b_n$ by convolving a raw
mini-XV diagram formed from all slices but the slice of interest. Then this
beam smearing component is subtracted from the observed slice, and the
normalisation $b_0$ is applied. The raw un-beam-smeared mini-XV diagram is
built using Eqn.~(\ref{ch4-vel-profile}). This requires the intrinsic
kinematics which are only known for radii at $R > R^{\prime}$
determined from fits outside the slice of interest. To estimate both
parts of the mini-XV diagram, we conduct an initial radial decomposition fit
across the whole observed XV diagram.  This first loop fit is 
accurate, except near
the galaxy centre, where the high gradients of rotation and velocity
dispersion over $R^{\prime}$ distort the line profiles. The kinematics
near the galaxy centre are estimated (for beam smearing correction) by
extrapolating inwards using the higher quality fits over the rest of the galaxy.
Subsequent iterations of the radial decomposition fitting process can
then be beam corrected using the mini-XV diagram method, whereby the mini-XV
 diagram is formed using these estimated kinematics. 

This beam correction is essentially a deconvolution of the XV map to
remove the effect of the telescope beam. As such the sampling of the
beam, i.e. the number of pixels per beam, affects the success of the
deconvolution, as the deconvolution process amplifies the noise. 
The recommended sampling
for deconvolution is 3 pixels per beam FWHM$_{\theta}$. To recover a slightly
higher spatial resolution, which is useful when modelling the more distant
galaxies, we used a beam sampling of 5 pixels when forming the \HI\
image cubes. Both sampling densities work well, however a yet higher
sampling density ($\gesim$8) dramatically reduces the peak
signal-to-noise of the beam corrected XV slice, as the telescope beam
naturally limits the spatial resolution of the observations. As the
galaxies in our sample lie at recession velocities from $400-2500$
\kms, and distances of $3-30$ Mpc, the FWHM$_{\theta}$ beam size of the galaxy
observations is $0.1-1.0$ kpc.  Given that the kinematics can be
recovered on a scale of $1/5$ of the beam FWHM$_{\theta}$, this leads to a
recovered spatial resolution of $20-200$ pc for our galaxies. 

The beam correction is necessary for galaxies with large beam sizes
$\gesim 500$ pc, as for these galaxies the beam smearing is
sufficient to distort the slice shape away from a sum of Gaussians,
over the major axis range where the kinematics vary significantly over
the span of the beam.  Without beam correction, radial
decomposition fitting of the inner disk will tend to oscillate 
with a large scatter, 
due to the intrinsicly non-multi-Gaussian shape of the velocity 
profiles in these 
sightlines. This occurs as an erroneous fit to gas on the line 
of nodes in any one 
sightline will cause the fit to the next inwards radius to 
overcompensate, resulting 
in an oscillating fit over the beam-smeared domain.

\begin{subfigures}
%
%
\begin{figure*}[t]
  \centering
\includegraphics[width=15cm]{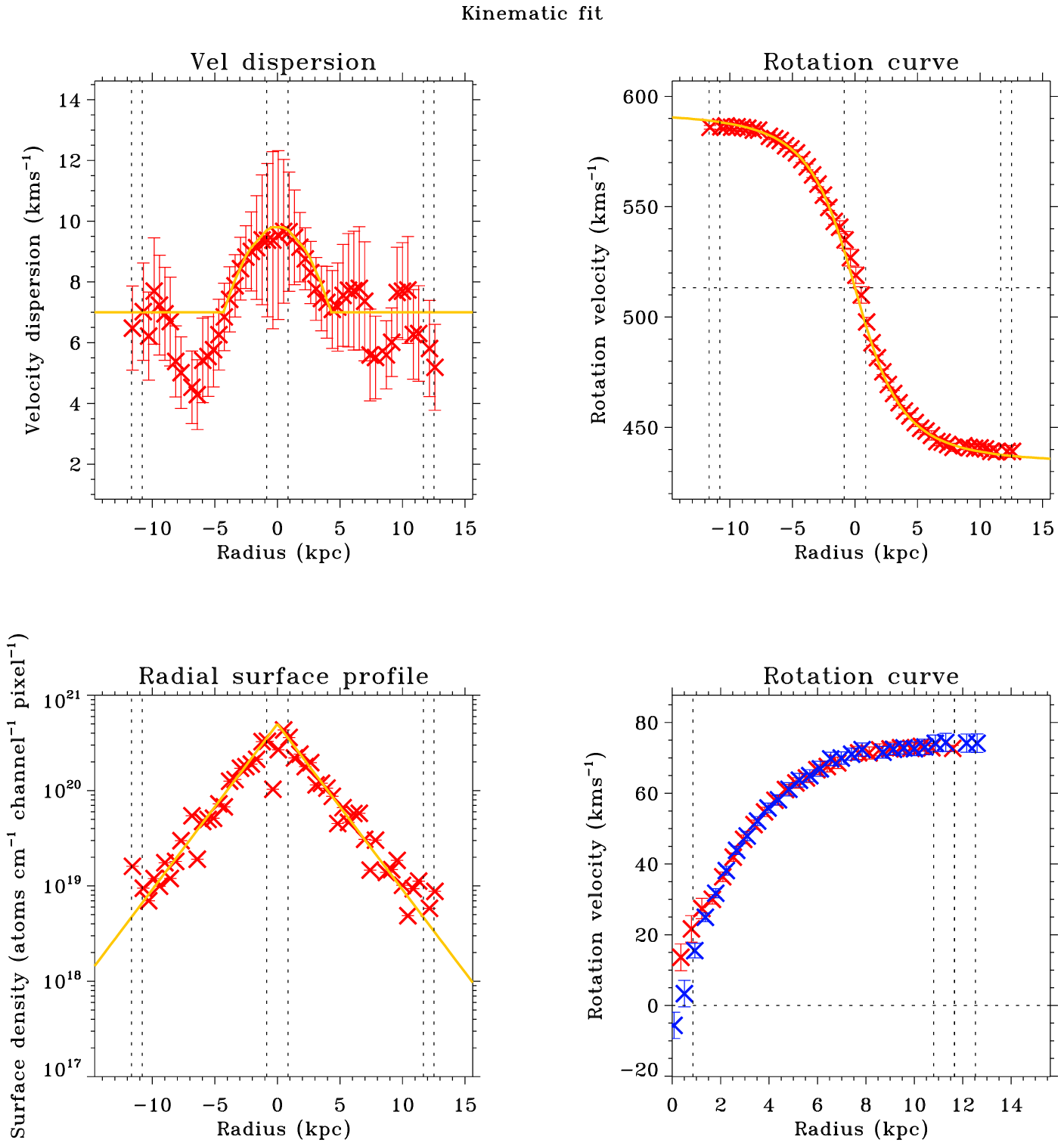}
  \caption[Measured kinematics of an XV diagram observed with a
  FWHM$_{\theta}$= 2 kpc telescope beam and to a sensitivity
  $S/N_{max}$ of 80.]{Fitted kinematics (with beam correction)
    measured from XV diagram observed with a telescope beam of
    FWHM$_{\theta}$= 2 kpc and with a peak signal-to-noise of 80. To
    reduce the scatter in the measured kinematics, the velocity
    dispersion and rotation curve were averaged (post fitting) over an
    evenly weighted bin of width of 3 pixels. In
    Fig.~\ref{fig:beamcorr_xv_2.0beam_SN80}, the XV diagram built with the
    best-fit measured kinematics and observed to the same telescope
    specifications is compared to the original XV diagram.}
  \label{fig:beamcorr_kfit_2.0beam_SN80}
\end{figure*}

\setcounter{figure}{1}
\begin{figure*}[t]
  \centering
\includegraphics[width=15cm]{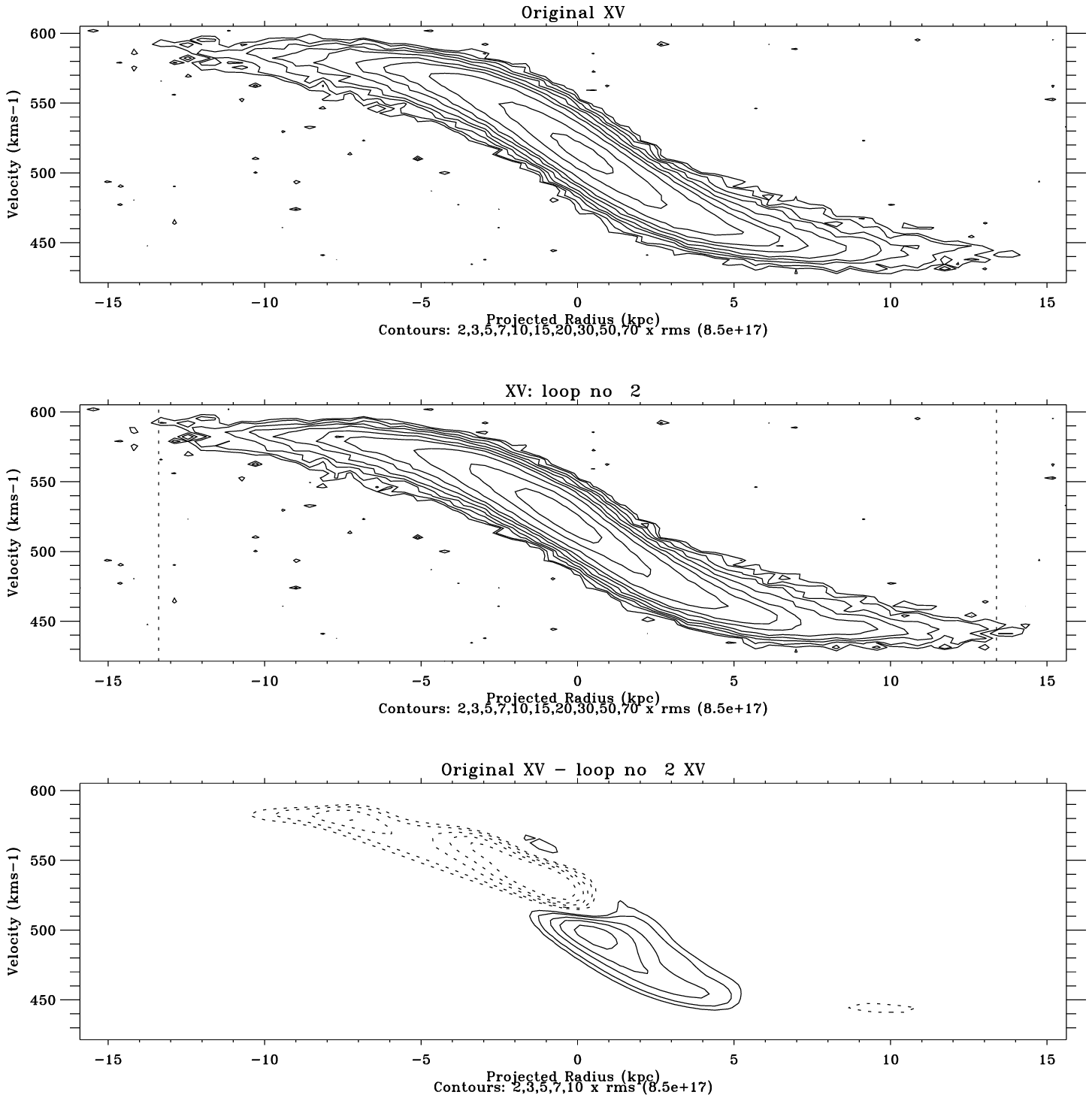}
  \caption[XV diagram (observed with a FWHM$_{\theta}$= 2 kpc telescope beam
  and peak $S/N$ of 80), XV diagram built with the measured kinematics and the
  residual after model XV diagram has been subtracted from the observed XV 
diagram.]{Top:
    Original XV diagram observed with a FWHM$_{\theta}$= 2 kpc telescope beam
    and peak $S/N$ of 80. Middle: The XV diagram built using the best-fit
    measured kinematics and observed with the same telescope
    resolution and sensitivity limits. The measured kinematics used to
    build the model XV plot are shown in
    Fig.~\ref{fig:beamcorr_kfit_2.0beam_SN80}. 
Bottom: The difference between the top and middle diagrams. Contour units are
2, 3, 5, 7, 10, 15, 20, 30, 50, 70 times the r.m.s. noise in the
channel maps.}
  \label{fig:beamcorr_xv_2.0beam_SN80}
\end{figure*}
\end{subfigures}

\begin{subfigures}
\begin{figure*}[t]
  \centering
\includegraphics[width=15cm]{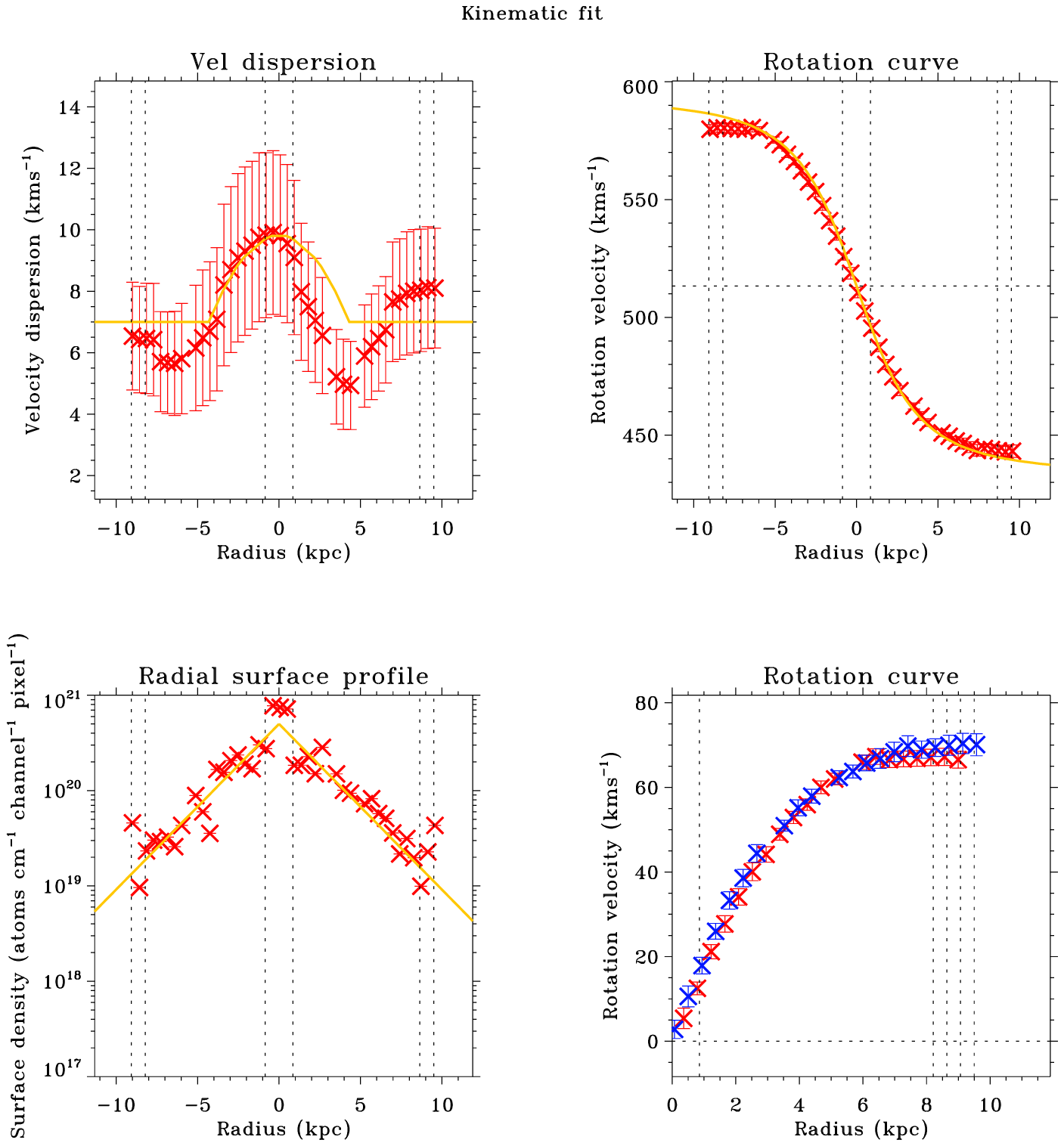}
  \caption[Measured kinematics of an XV diagram observed with a
  FWHM$_{\theta}$= 2 kpc telescope beam and to a sensitivity
  $S/N_{max}$ of 30.]{Fitted kinematics (with beam correction)
    measured from XV diagram observed with a telescope beam of
    FWHM$_{\theta}$= 2 kpc and with a peak signal-to-noise of 30. To
    reduce the scatter in the measured kinematics, the velocity
    dispersion and rotation curve were averaged (post fitting) over an
    evenly weighted bin of width of 9 pixels. In
    Fig.~\ref{fig:beamcorr_xv_2.0beam_SN30}, the XV diagram built with the
    best-fit measured kinematics and observed to the same telescope
    specifications is compared to the original XV diagram. The plots show that
    the larger beam of this observation allows the shape of the
    velocity dispersion to be recovered more accurately than a higher
    resolution observation with a $1.0$ kpc telescope beam, observed
    to the same noise limit (compare to
    Fig.~\ref{fig:beamcorr_kfit_1.0beam_SN30}).}
  \label{fig:beamcorr_kfit_2.0beam_SN30}
\end{figure*}

\setcounter{figure}{1}
\begin{figure*}[t]
  \centering
\includegraphics[width=15cm]{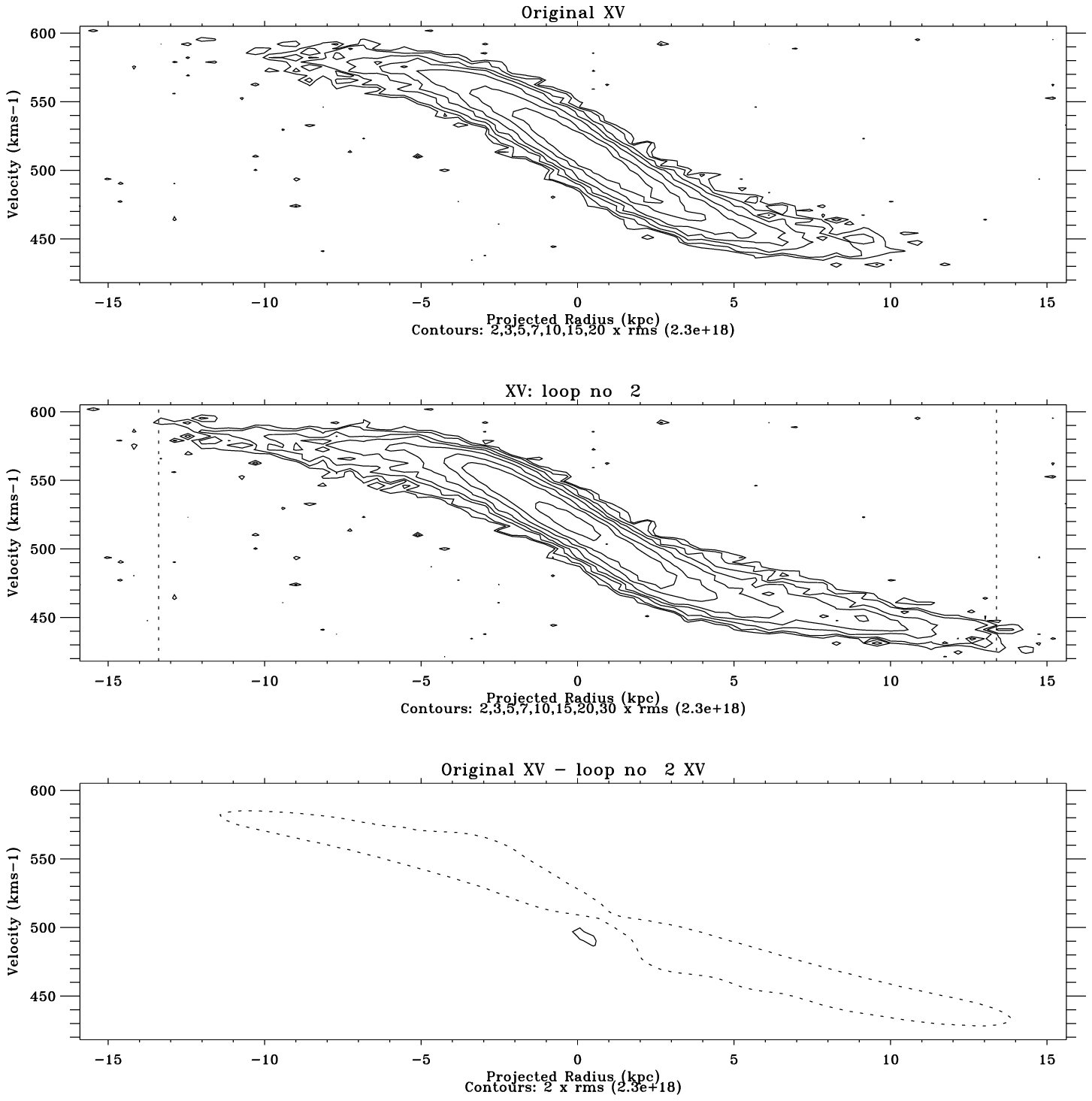}
  \caption[XV diagram (observed with a FWHM$_{\theta}$= 2 kpc telescope beam
  and peak  of 30) and XV diagram built with the measured kinematics.]{Top:
    Original XV diagram observed with a FWHM$_{\theta}$= 2 kpc telescope beam
    and peak $S/N$ of 30. Bottom: The XV diagram built using the best-fit
    measured kinematics and observed with the same telescope
    resolution and sensitivity limits. The measured kinematics used to
    build the bottom plot are shown in
    Fig.~\ref{fig:beamcorr_kfit_2.0beam_SN30}. Bottom: 
The difference between the top and middle diagrams. Contour units are
2, 3, 5, 7, 10, 15, 20, 30 times the r.m.s. noise in the
channel maps.}
  \label{fig:beamcorr_xv_2.0beam_SN30}
\end{figure*}
\end{subfigures}

\subsection{Iterating to ensure self-consistent fits}
\label{sec:iter-process}

To assess and improve the quality of the fitted kinematics, 
the process is iterated
by building a model galaxy from the original fitted kinematics,
simulating the observations and comparing the fitted properties of the
model $f_n[v(R),\sigma_{v,\HI}(R),\Sigma_{\HI}(R)]$ galaxy XV diagram to the
original fitted kinematics
$f_1[v(R),\sigma_{v,\HI}(R),\Sigma_{\HI}(R)]$. A schematic showing the
method is shown in Fig.~\ref{fig:ch4-iter-flowchart}. In the initial
loop, $loop_0$, the observed XV map is fitted to obtain a rough kinematic
model that is used to perform beam correction. This mini-XV diagram is then
used in the next fit to the observed XV map, which we will call $loop_1$. The
radial functions of \HI\ surface density, rotation and velocity 
dispersion fitted in 
$loop_1$ fit, which are the first fits incorporating beam correction, 
are then used as
the first kinematic model for a new XV diagram. By iterating the fitting 
process, 
and incrementing the kinematic model between fits, the process is repeated 
until the fitted kinematics from the model equal the original
fitted kinematics with beam correction to within a critical threshold.

By intrinsically incorporating fitting of the gas velocity dispersion
into the radial decomposition method, it is possible to replicate the observed
XV diagram from the best fit kinematics as all three functions characterising
the XV map are obtained: rotation curve $v(R)$, \HI\ velocity
dispersion $\sigma_{v,\HI}(R)$ and the deprojected \HI\ surface density
$\Sigma_{\HI}(R)$. Unlike earlier methods of measuring rotation curves or
\HI\ velocity dispersion, this method is internally self-consistent.

\section{Quality testing on synthetic galaxy observations}
\label{sec:quality-test}

The quality of the derived kinematics is quantified in two ways.
Firstly, we calculate the total absolute difference between the best fit and
original fit for each of the three functions: $v(R)$,
$\sigma_{v,\HI}(R)$ and $\Sigma_{\HI}(R)$. This difference is
summed over all measured slices across the galaxy and divided by the
number of measured slices to present a average difference at each radius.

Secondly, the XV map formed from the best fit kinematics is compared
to the observed XV map by computing the residual. The residual map 
(with absolute values for the residuals) is
normalised by the observed XV map to obtain a total percentage
difference between the actual galaxy XV diagram and the best fit model. The
residual XV map is a good indicator of surface density and rotation
curve accuracy, but is relatively insensitive to errors in the
measured velocity dispersion.

\subsection{Effect of noise}
\label{sec:ch4-disc-eff-noise}

The noise level is of most concern when fitting observations that suffer
significant beam-smearing, as the beam correction method uses
deconvolution, which is particularly sensitive to noise. To isolate 
the effect of noise from that due to beam-smearing, we
tested the radial decomposition method on high spatial resolution XV
maps (observed with FWHM$_{\theta}$= 300 pc) with different
sensitivities corresponding to peak signal-to-noise ratios ranging
from 20 to 80. In each case the same galaxy model is used, featuring a
differential rotation curve, an exponential face-on \HI\ surface
density (with a \HI\ scalelength of $2.5$ kpc) and a flat outer \HI\
velocity dispersion that rises parabolicly in the inner disk from an
outer plateau of $7$ \kms\ to a central peak of $10$ \kms. In 
Fig.~\ref{fig:xv_raw_obs-diff_SN} we show the XV diagrams
for this synthetic galaxy without noise and with noise introduced
at peak $S/N$ levels of 80 and 20. The upper diagrams, labelled
`raw unedged', are the pure synthetic diagrams that would be observed 
in the absence of noise. The lower ones show the effects of introducing noise.
The radial (horizontal) scales in the diagrams left and right
are different and have been chosen such that it the observable radial extents
fit within the frames. This shows the severe effects of noise.

In Figs.~\ref{fig:smallbeam_kfit_SN80}
and \ref{fig:smallbeam_kfit_SN20}, we show the measured kinematics
from the simulated observation of this model galaxy with noise levels
corresponding to peak
signal-to-noise ratios of $80$  and $20$. In each figure the actual
kinematics are shown in grey (yellow).
As mentioned above, both the rotation curve and the
radial surface density are well measured, even in the low sensitivity
$S/N$= 20 and $S/N$= 30 figures. However, the \HI\ velocity dispersion is
particularly sensitive to high noise levels that submerge the outer
\HI\ distribution. Comparison of an ideal XV map and XV map generated
with finite telescope sensitivity and resolution
(see Fig.~\ref{fig:xv_raw_obs-diff_SN}) demonstrates the limited
\HI\ detected at low sensitivities. But despite the low flux recovery,
these simulations show that low sensitivity observations still yield
reasonably accurate kinematics, albeit with dramatically increasing
measurement error at lower sensitivities.

To reduce the scatter, an evenly weighted running mean was applied to
the measured fits after the XV diagram was fitted with the radial
decomposition method. By averaging the measured kinematics every 3
or more pixels in radius, it was possible to reduce the scatter in
measured velocity dispersion and recover the radial trend of the kinematics 
with good confidence. This proved to be more revealing, although it 
effectively lowered the spatial resolutions of the measurement. The 
simulations show that measurements of galaxies observed with peak $S/N <
100$ benefit from this post-fit averaging. Given an XV image
gridded with a beam sampling density of 5 (i.e. each pixel is $1/5$ of
the beam FWHM$_{\theta}$) and beamwidth FWHM$_{\theta}$=
300 pc, the final post-averaging resolution is still high at $180$ or
$300$ pc after averaging over a $3$-pixel or $5$-pixel wide bin. A
post fit averaging interval of 5 was used for the $S/N_{max}$= 80
dataset; while this was increased to 7 for the noisy
$S/N_{max}$= 20 XV map. As shown by the plots in
Figs.~\ref{fig:smallbeam_kfit_SN80}
and \ref{fig:smallbeam_kfit_SN20}, it was possible to recover the
radial trend in the velocity dispersion for XV diagrams with these 
kinematics, providing the peak $S/N > 20$.

In Figs.~\ref{fig:smallbeam_kfit_SN80}
and \ref{fig:smallbeam_kfit_SN20}, we show the measured kinematics 
from the simulated observation of this model galaxy with noise levels 
corresponding to peak
signal-to-noise ratios of $80$  and $20$. In each figure the actual
kinematics are shown in grey (yellow).
As mentioned above, both the rotation curve and the
radial surface density are well measured, even in the low sensitivity
$S/N$= 20 and $S/N$= 30 figures. However, the \HI\ velocity dispersion is
particularly sensitive to high noise levels that submerge the outer
\HI\ distribution. Comparison of an ideal XV map and XV map generated 
with finite telescope sensitivity and resolution 
(see Fig.~\ref{fig:xv_raw_obs-diff_SN}) demonstrates the limited 
\HI\ detected at low sensitivities. But despite the low flux recovery, 
these simulations show that low sensitivity observations still yield 
reasonably accurate kinematics, albeit with dramatically increasing 
measurement error at lower sensitivities.

In the slice fits of the edge of the galaxy, the measurements are 
biased by two effects.
Firstly, in the outer few slices the radial decomposition method
models all the flux as originating from a single radial bin. 
Consequently, the measured
velocity dispersion is broadened by the projected rotation of gas
away from the line-of-nodes, i.e. at radii higher than $R$=$R^{\prime}$;
similarly the measured surface density is overestimated due to the
attribution of flux away from the lines-of-nodes. This problem is 
systematic of the radial decomposition method, although it is 
relatively unimportant
as it only effects the outer few slices.

Secondly, the fits to the velocity profile of slices with very low
peak $S/N$ ($\lesim 5-8$) are biased by noise peaks dominating the
underlying flux profile. The radial decomposition fitting algorithm
starts fitting where it finds velocity profile slices with flux
greater than $2\sigma$ in 4 adjacent velocity channels. A galaxy
with an exponentially declining radial surface density typically
has a peak $S/N$ $\lesim 5-8$ in the slices at  the outermost $1-2$ kpc.
Gaussian fitting of these low $S/N$ velocity profiles is particularly
sensitive to underestimating the velocity dispersion, due to fitting
the noise peaks instead of the underlying flux profile. As noise
peaks typically have a FWHM of 1-2 velocity channels, the velocity
dispersion measurements in these outer few kpc is contaminated with
noise fits exhibiting a velocity dispersion of $\sigma_v \sim
(1-2\,\textrm{channels}) / (2 \sqrt{2 \log 2} ) \sim 1-2$ \kms. As
the radial decomposition fitting method progressively fits inwards
using the outer fits to model the gas at radii away from the line-of-nodes
in each velocity profile, erroneous fits cause subsequent measurements
to oscillate around the correct value.  These erroneous fits only
effect the adjacent few slice fits as the typically near-exponential
radial surface density in the outer \HI\ disk causes the flux
contribution of outer radii to be relatively small.

In summary, the radial decomposition method is able to successfully measure
the radial distributions of the velocity dispersion, rotation
velocity and surface density for XV diagram observations with peak $S/N$ greater
than 20. This could perhaps be undertaken with lower $S/N$ maps, if the
observation is actually detecting the bulk of the actual \HI\
distribution, as could be the case on a galaxy with an intrinsically
shallow radial surface density. The method breaks down for lower $S/N$
maps which are dominated by the noise over a larger major axis extent
of the detected gas distribution. Also, as can be seen from the
above figures, the major limitation of the low signal-to-noise
observation of the above galaxy model is that one is only detecting
the inner half of the gas disk, because the outer disk \HI\ lies
below the $2\sigma$ noise level.  The outermost radius with sufficient
flux above $2\sigma$ to allow kinematic fitting is marked by vertical
dashed lines on the XV maps observed with different noise limits
in Fig.~\ref{fig:xv_raw_obs-diff_SN}.

\subsection{Effect of spectral resolution}
\label{sec:ch4-disc-eff-instr-broad}

Modern radio telescope array correlators of the ATCA and VLA have a
spectral resolution of $3-5$ \kms, when configured for a frequency
bandwidth sufficient to cover the rotational velocity range of a
small-to-medium spiral galaxy ($v_{max} \lesim 200$ \kms). This is
sufficient to resolve the FWHM due to the typical intrinsic gas velocity
dispersion of $7$ \kms\ (the commonly adopted value) by $3-5$
channels. For this reason we did not investigate how
spectral resolution changes affected the measurement accuracy, 
although it should probably be considered if low \HI\ velocity 
dispersions are found. All measurements of simulated \HI\ observations 
were corrected for instrument broadening, according to
Eqn.~(\ref{eq:ch4-instr-broadening-corr}).

\begin{subfigures}
\begin{figure*}[t]
 \centering
\includegraphics[width=11cm]{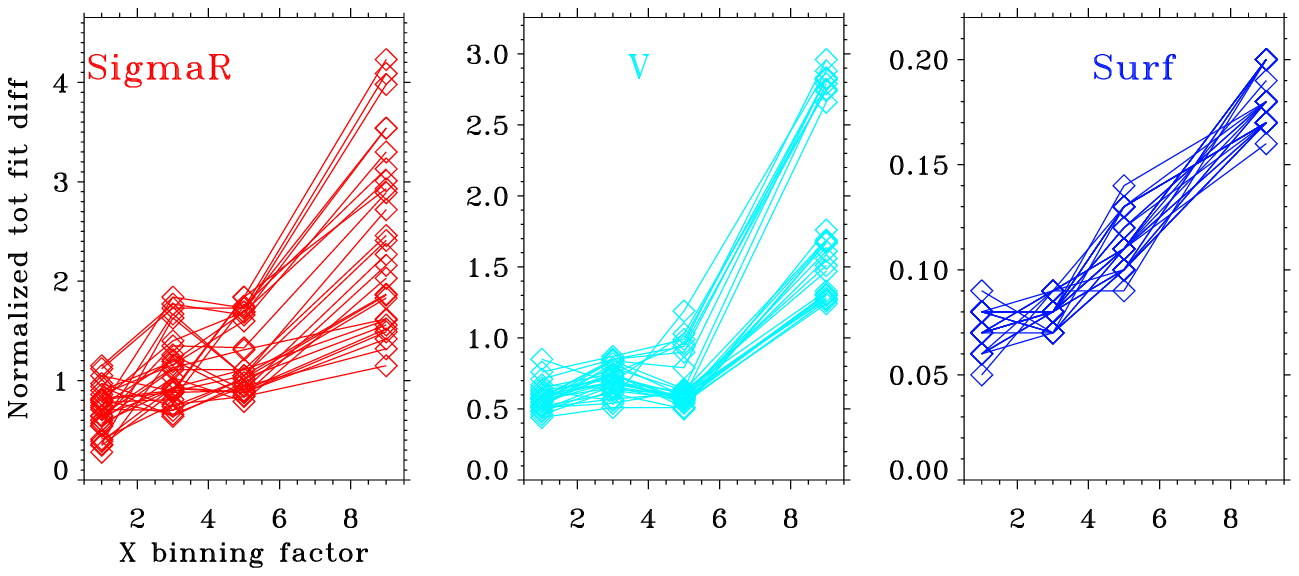}
\includegraphics[width=4.2cm]{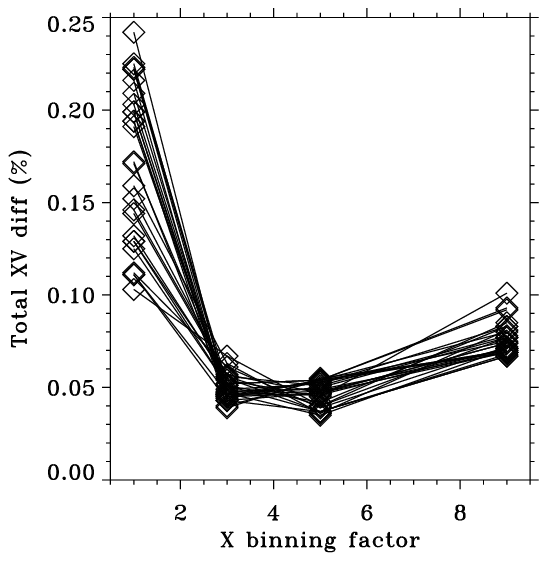}
\includegraphics[width=11cm]{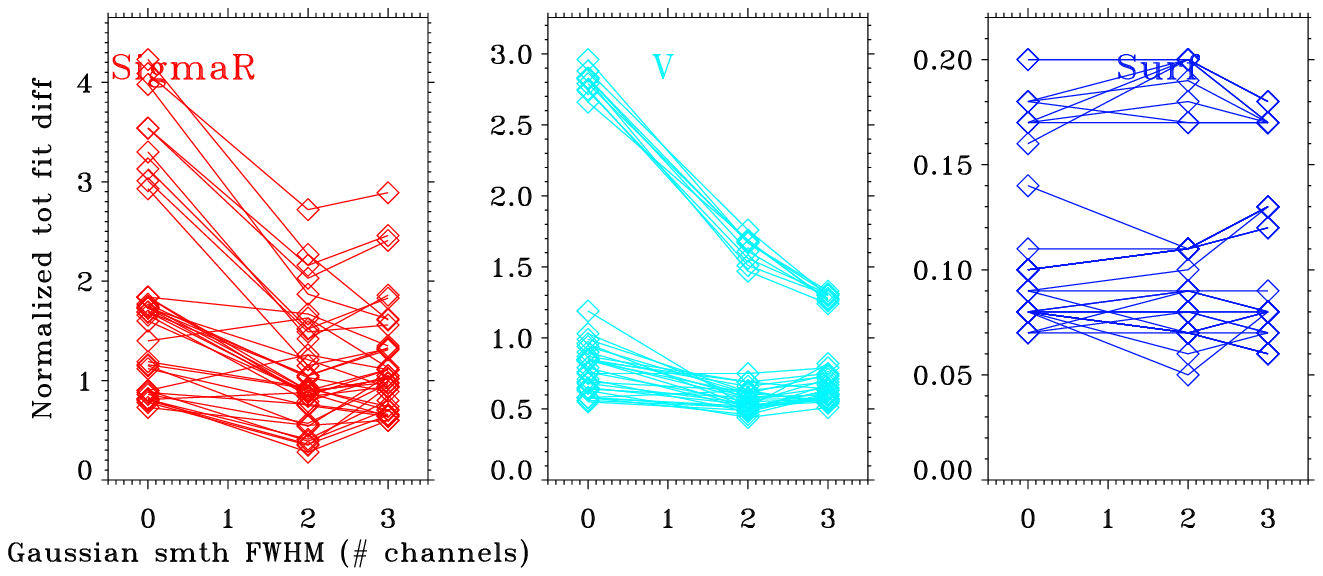}
\includegraphics[width=4.2cm]{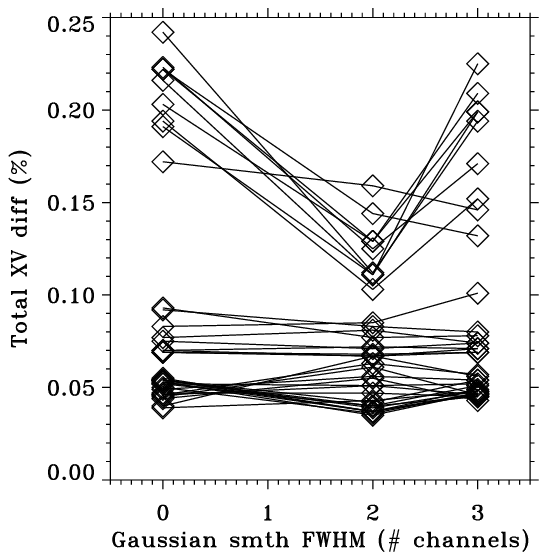}
\includegraphics[width=11cm]{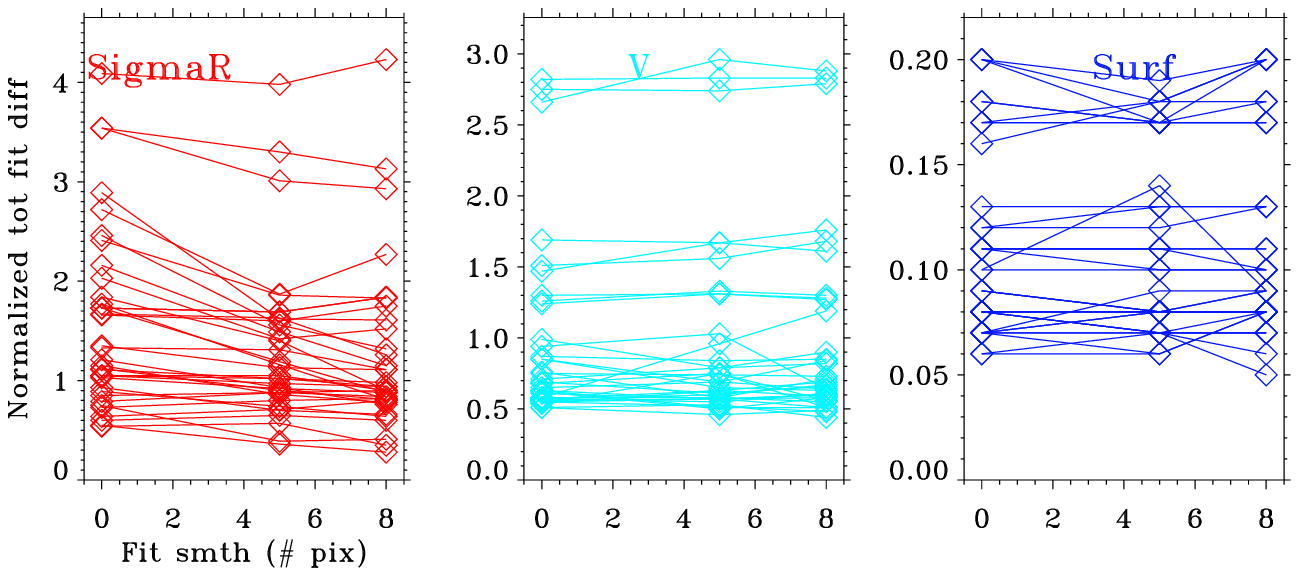}
\includegraphics[width=4.2cm]{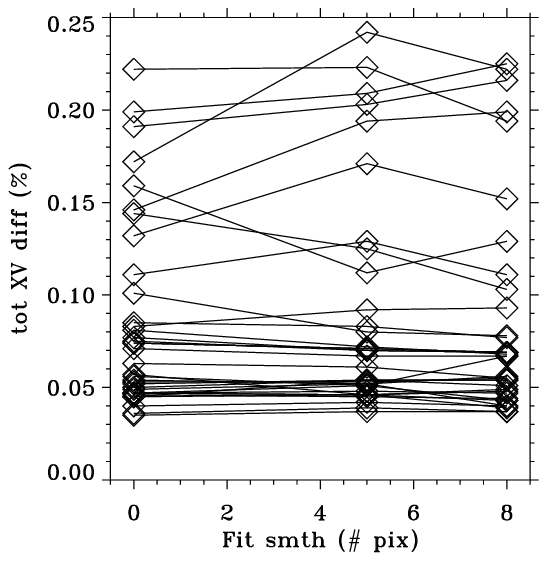}
\includegraphics[width=11cm]{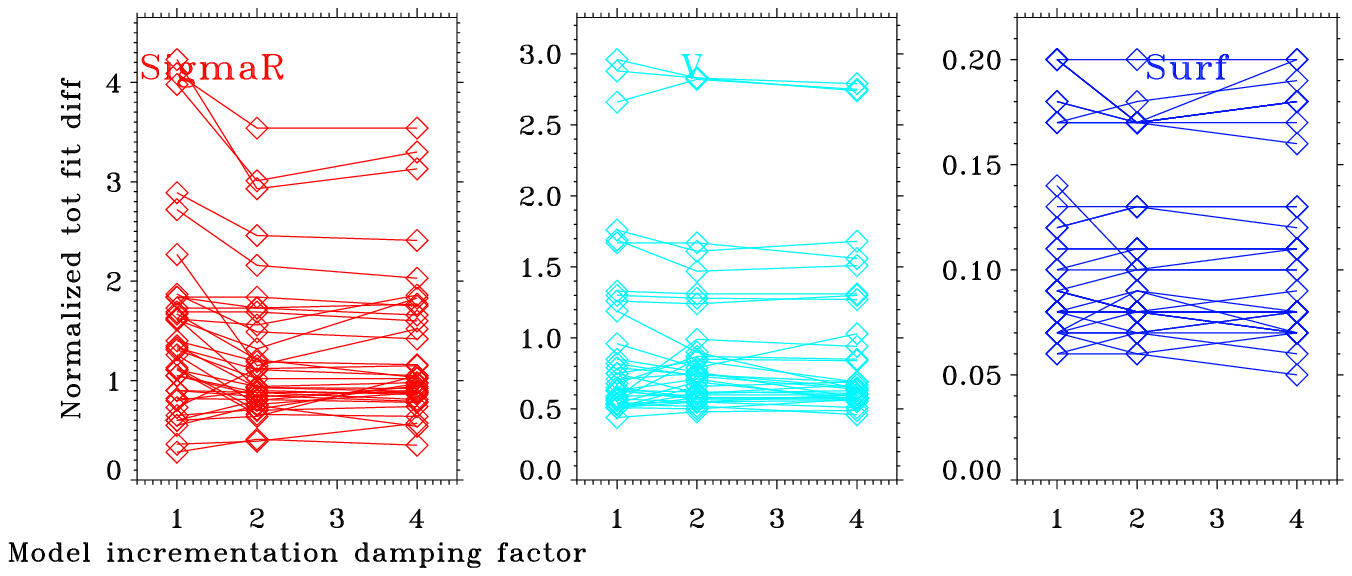}
\includegraphics[width=4.2cm]{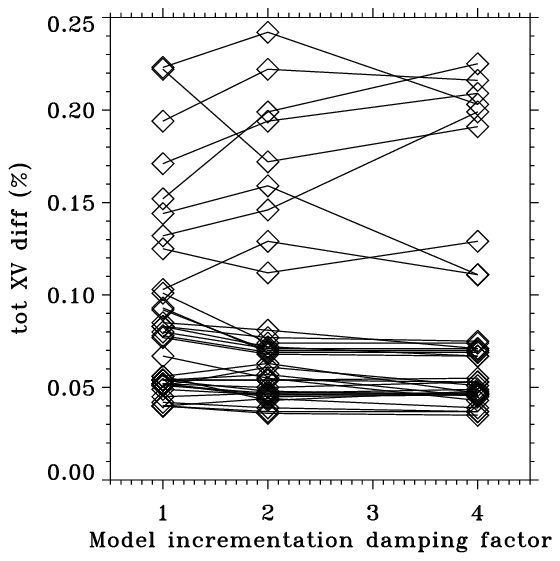}
  \caption[Effect of different alterations to the radial decomposition
  method on the measured kinematics of a synthetic galaxy with
  signal-to-noise $SN$= 80, and a small beam $\theta_{syn}$= 0.3
  kpc]{These plots display measures of the fit to the kinematics and
    how these are affected by varying the iterative onion peeler for a
    synthetic galaxy with signal-to-noise $S/N$= 80, and a small beam
    $\theta_{syn}$= 0.3 kpc.  The first three columns show the integrated
    absolute difference between the final model and the 
    the original XV diagram for each kinematic function (from left to
right respectively velocity dispersion `SigmaR', rotation velocity 
`V' and surface density `Surf'), normalised by the
    number of sightlines within a fit. The units for the total fit
    difference of the velocity dispersion and the rotation curve are
    in \kms, while the surface density difference is measured in log
    flux. The fourth column displays the integrated flux difference
    between the final XV diagram built from the best fit kinematics, and the
    original XV diagram, as a percentage of that in the original XV diagram. 
See the text for further explanation.}
  \label{fig:fv_SN80_0.3beam}
\end{figure*}

\setcounter{figure}{1}
\begin{figure*}[t]
 \centering
\includegraphics[width=11cm]{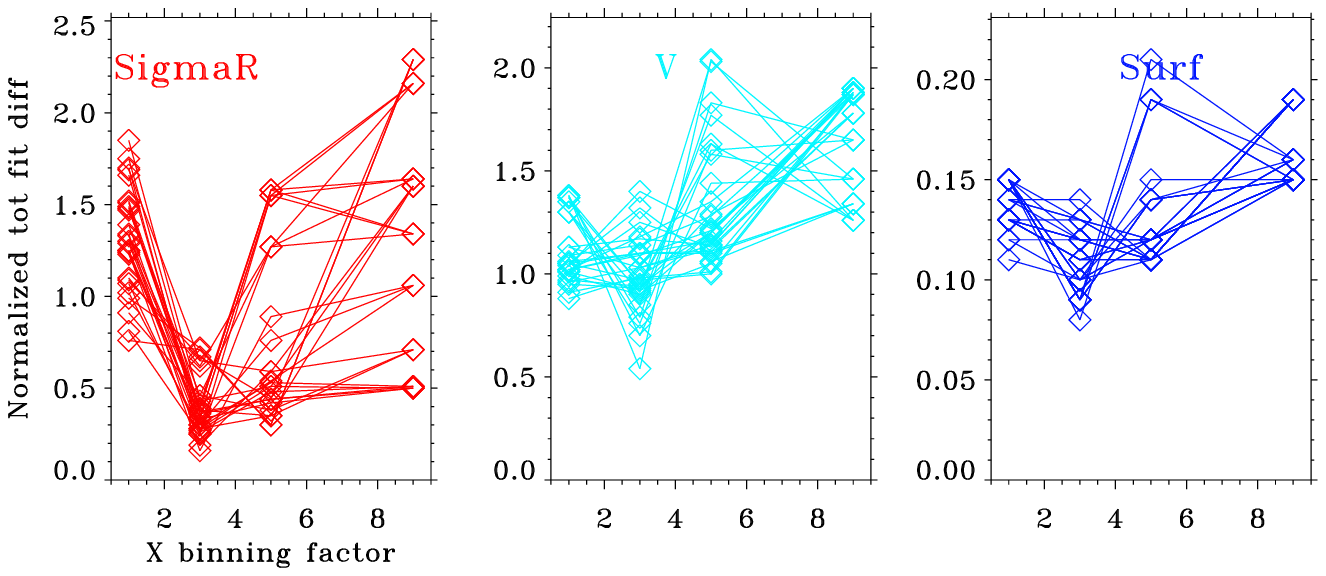}
\includegraphics[width=4.2cm]{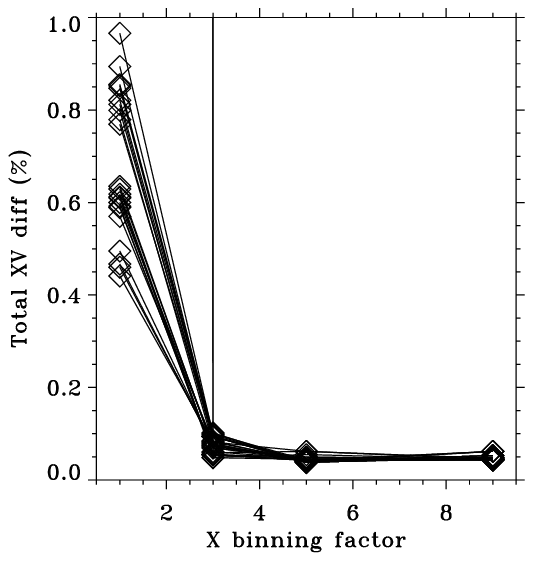}
\includegraphics[width=11cm]{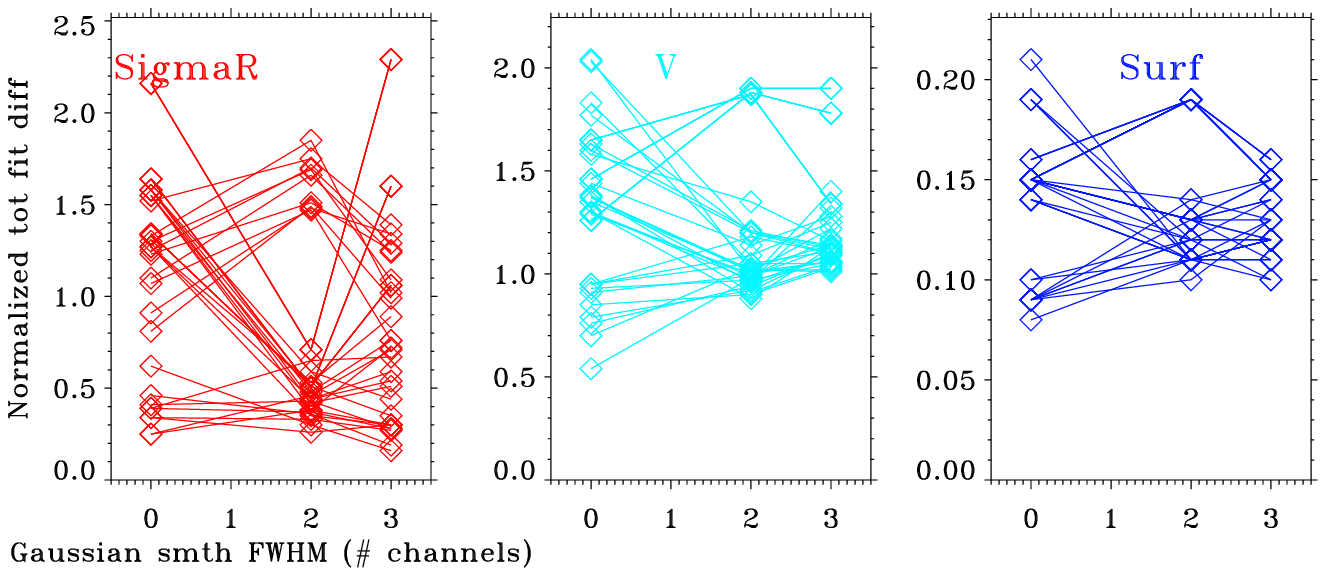}
\includegraphics[width=4.2cm]{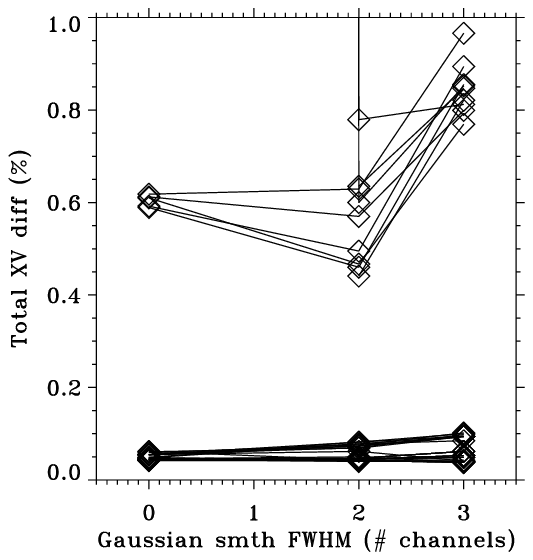}
\includegraphics[width=11cm]{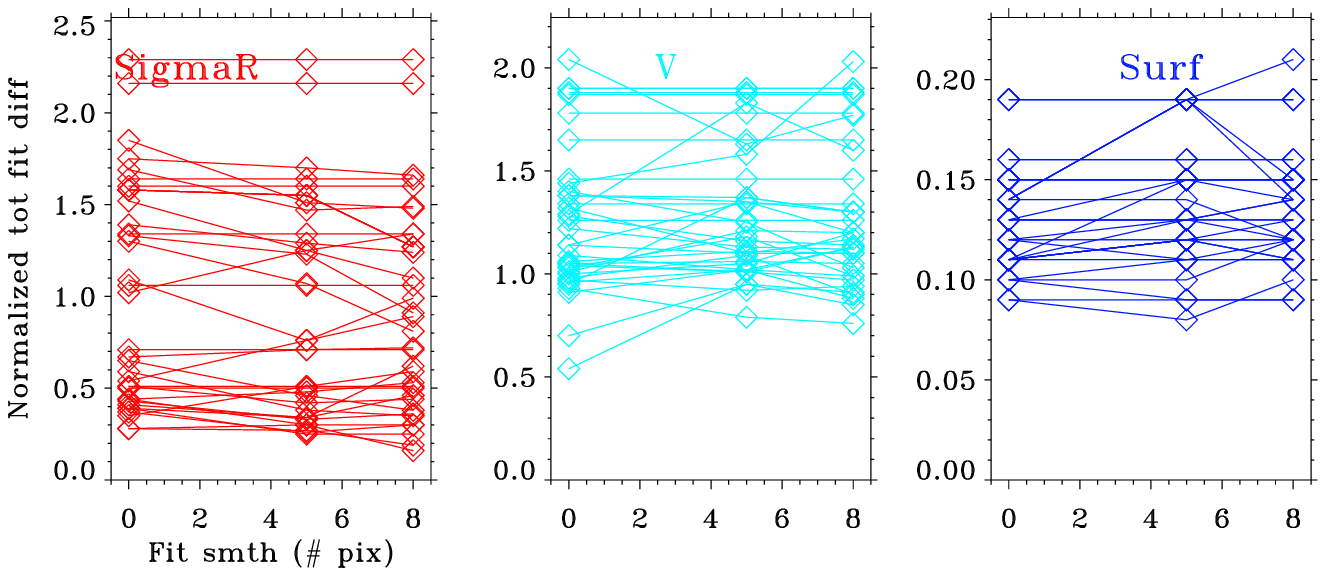}
\includegraphics[width=4.2cm]{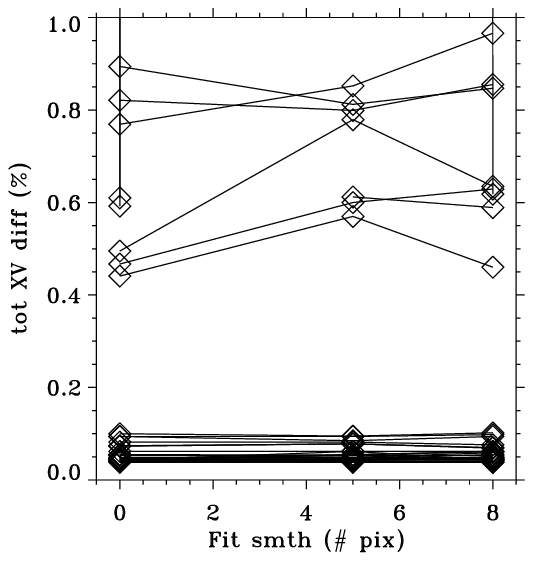}
\includegraphics[width=11cm]{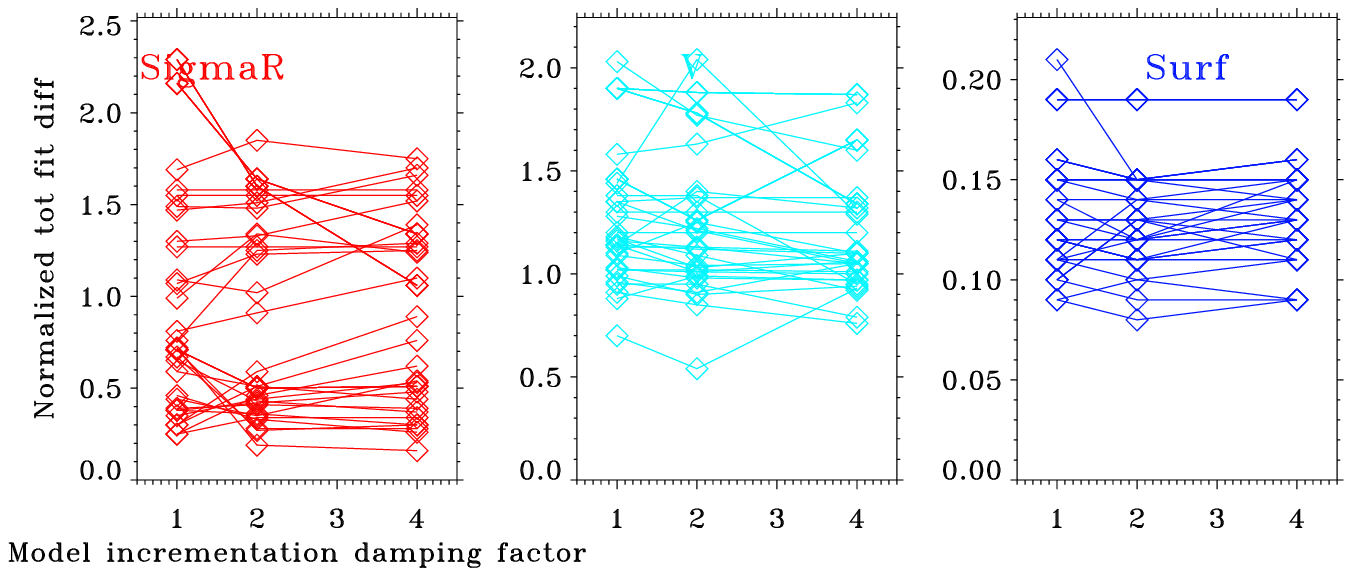}
\includegraphics[width=4.2cm]{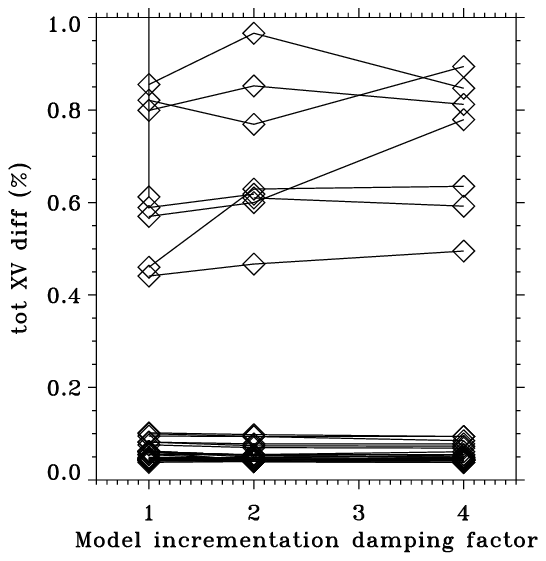}
  \caption[Effect of different alterations to the radial decomposition
  method on the measured kinematics of a synthetic galaxy with
  signal-to-noise $SN$= 30, and a small beam $\theta_{syn}$= 0.3
  kpc]{These plots display the same measures of the fit to the kinematics
    as in Fig.~\ref{fig:fv_SN80_0.3beam}, but now 
    for a synthetic galaxy with signal-to-noise $S/N$ =30.
   Lay-out and units are the same as in
    Fig.~\ref{fig:fv_SN80_0.3beam}.}
  \label{fig:fv_SN30_0.3beam}
\end{figure*}
\end{subfigures}

\subsection{Effect of spatial resolution}
\label{sec:ch4-disc-eff-beam}

The shape of the beam is the largest factor apart from the sensitivity 
affecting 
the accuracy of the measured kinematics. But the effect of the beam smearing 
on the measured \HI\ kinematics is only serious if the intrinsic galaxy 
kinematics 
has a steep gradient over the scale of the beam. As a result, the observed
\HI\ distribution is more distorted over projected radial domains with steep
gradients of rotation, velocity dispersion or surface density than those 
parts of 
the galaxy with shallower gradients. In practice, the observed \HI\ 
distribution
of a small galaxy with a maximum rotation speed of $\lesim 100$ \kms\ 
and a slowly rising inner velocity dispersion is only minimally
distorted by a telescope beam with FWHM$_{\theta}$= 300 pc. (See
Figs.~\ref{fig:beameff_kfit_1.0beam} and \ref{fig:beameff_kfit_2.0beam}
to view the magnitude of the
beam-smearing effects of different the beam size on the measured
kinematics of the model galaxy.)

Calculation of the dark halo mass density requires smooth functions
of the kinematics. Consequently, it is more important to determine 
the gas kinematics on scales of
$1$ kpc, than $100$ pc. The sensitivity of the observations and accuracy 
of the radial kinematics was improved by convolving the observations by 
a larger beam size prior to fitting. 
This is demonstrated in Figs.~\ref{fig:beamcorr_kfit_1.0beam_SN30}
and \ref{fig:beamcorr_kfit_2.0beam_SN30} which show the measured kinematics
using beam correction from the same galaxy observed with
FWHM beams of $1$ and $2$ kpc, respectively, at a low sensitivity with
peak $S/N$ of $30$.

However, as the beam-smearing correction is performed by deconvolution
--- which amplifies the effective noise --- it is important to perform
beam correction only over the major axis range affected by beam
smearing. Otherwise one is adding unnecessary scatter to the
measurements across the whole radial extent of the galaxy. To
investigate the effect of beam-smearing on the measurements kinematics, 
simulations we analysed simulated, noise-less XV maps.
Figs.~\ref{fig:beameff_kfit_1.0beam} and \ref{fig:beameff_kfit_2.0beam}
show the results of fits when beam-smeared with a telescope beam
with  an FWHM$_{\theta}$ of $1.0$ and 2.0 kpc beams.  The beam-smearing
effect becomes quite dramatic in the measurements of galaxies
observed with large beams (2 kpc or more). 
Because of the location of the grid points,
the uncorrected kinematics shows some marked asymmetries (e.g. 
Fig.~\ref{fig:beameff_kfit_1.0beam})
which are almost completely corrected by the beam-smearing correction
process (Figs.~\ref{fig:beamcorr_kfit_1.0beam_SN80} and 
\ref{fig:beamcorr_kfit_1.0beam_SN30}).

As explained in Sect.~\ref{sec:beam-corr}, beam-smearing causes the
velocity profile of a slice to deviate from a sum of Gaussians. 
As a result, the measured fits
of the velocity dispersion and rotation velocity oscillate around the
true value, due to the progressive nature of the fitting
process. Similar oscillations also occur in the fitted surface density; however 
these oscillations are on flux scales lower than the scatter due to noise.

The best procedure is to first perform an
initial radial decomposition fit of the galaxy without beam-smearing
correction. This gives accurate kinematic fits at radii where there
is no large intrinsic gradients of the kinematics. The radii affected
by beam-smearing are determined from the radii with oscillating measured fits. 
Subsequent iterations of the fitting process, are performed using beam smearing 
in this radial range. As
explained in Sect.~\ref{sec:beam-corr}, the shape of the inner
galaxy kinematics are estimated by extrapolating inwards to build the
mini-XV diagram used in beam-correcting the radii affected by beam-smearing.
Fortuitously, the large radial gradients of the intrinsic kinematics 
that necessitate beam smearing correction occur predominantly in the 
inner disk. Most late-type, gas-rich galaxies contain sufficient \HI\ 
in the inner disk to apply the beam smearing correction method. 

Figs.~\ref{fig:beamcorr_kfit_1.0beam_SN80}
to \ref{fig:beamcorr_kfit_2.0beam_SN30} show the measured
kinematics fits obtained with beam-smearing correction of XV maps
obtained from moderate (FWHM$_{\theta}$= 1.0 kpc) and low spatial
resolution (FWHM$_{\theta}$= 2.0 kpc) XV maps. The measurements are
undertaken on both low and high sensitivity XV maps (peak $S/N$= 30 and
$S/N$= 80, respectively) at each spatial resolution. The XV maps built
with the measured kinematics are displayed in
Figs.~\ref{fig:beamcorr_xv_1.0beam_SN80} to
\ref{fig:beamcorr_xv_2.0beam_SN30}.

The residual effects of beam-smearing after
 correction in high S/N data (peak S/N = 80) can be appreciated by
 comparing Figs.~\ref{fig:smallbeam_kfit_SN80} (300pc beam), 
\ref{fig:beamcorr_kfit_1.0beam_SN80} (1 kpc beam) and 
\ref{fig:beamcorr_kfit_2.0beam_SN80} (2 kpc beam).
It can be seen that the beam correction process does a good job of
correcting the beam-smearing that occurs over the inner disk. The
velocity dispersion is well recovered over large radial scales, except
for the measurements on the low signal-to-noise ($S/N$= 30) dataset
observed with the $1$ kpc beam. Very high signal-to-noise observations
($S/N \gesim 100$) were able to recover the radial velocity dispersion
structure on sub-beam scales to accuracies of less than $1$ \kms,
however measurements to observations with large beams FWHM$_{\theta}
\ge 1$ kpc at $30 \lesim S/N \lesim 100$ are limited to recovered the
velocity dispersion on radial scales of $\sim$1 FWHM$_{\theta}$ and an
uncertainty of approximately $1-2$ \kms.

The measured rotation curve is accurately recovered. 
The surface density is also recovered with good accuracy,
though the uncertainty is more sensitive to the noise level of the XV
map.

\subsection{Fine tuning the iterative method}
\label{sec:ch4-disc-finetuning}

To minimise the noise sensitivity of the velocity dispersion fitting 
method, various adaptions were trialed on simulated observations with 
different spatial resolutions and noise levels. These 
adaptions trialed were:

$\bullet $ Binning of the $R^{\prime}$ dimension of the XV map 
prior to fitting to
  increase the signal-to-noise.

$\bullet $ Smoothing of each velocity profile slice through the XV map by a
  Gaussian of known dispersion prior to fitting kinematics to each
  velocity slice. Profile broadening by a Gaussian with varying FWHM from 
  3 to 6 velocity channels was tried. This additional known source of  velocity 
  dispersion was taken into account in the derivation of the \HI\ 
velocity dispersion.

$\bullet $ Smoothing of the derived fit $\sigma_{v,\HI}(R)$ with a Gaussian
  FWHM of varying number of $R^{\prime}$ pixels. The smoothed fits
  were then compared to determine the new model $\sigma_{v,\HI}(R)$.
  Only the velocity dispersion fit was smoothed, as the rotation curve
  and deprojected surface density fits exhibited a very low scatter.

$\bullet $ Damping of the model incrementations $\Delta$ (see
  Fig.~\ref{fig:ch4-iter-flowchart}) to minimise oscillations in the 
fits used to create the new model. Damping was only applied to the
  model incrementation of the velocity dispersion, as the small
  scatter of the fits to the rotation curve and \HI\ surface density
  removed the need for any model damping.

Several variations were tried for each adaption, resulting in 108
separate simulations, which were run on both a high ($S/N$= 80) and low
($S/N$= 30) signal-to-noise XV map observed with a $0.3$ kpc beam. The
galaxy model used was a exponential \HI\ surface density, constant \HI\
velocity dispersion and a differential rotation curve. 
Two indicators of the accuracy were measured.

Firstly, a residual XV diagram was calculated by subtracting the original XV
 diagram from the XV diagram 
formed from the best fit kinematics. The summed absolute residual
XV flux in \%\ of the total flux in the maps
is displayed in the plots in the far right column of
Fig.~\ref{fig:fv_SN80_0.3beam} for the $S/N$= 80 results, and
Fig.~\ref{fig:fv_SN30_0.3beam} for the $S/N$= 30 results. While the
residual XV diagram is a good indicator of measurement accuracy on large
scales, it is quite insensitive to measurement errors on small
radial scales.
Except for one fit which diverged in the simulations on low
signal-to-noise XV diagram, the summed residual was not more than $1.0\%$
($0.25\%$ in the SN80 dataset) of the total flux in the original map
for both the low and high $S/N$ measurements, being approximately three
times greater in the fit to the low signal-to-noise map.

Secondly, we also used the difference between the kinematics measured 
from the XV diagram formed with the best fit kinematics and the
kinematics measured from the observed XV diagram. The difference in fit 
(again absolute values) at each radius was then summed to form
a total difference for the fit to that kinematic function. We were able 
to compare the fits obtained from binned and un-binned XV maps, by 
normalising the total
difference in fit by the number of sightlines used in the fit. 
The normalised total fit difference for each kinematic function is 
displayed in the
plots in the first three columns of Fig.~\ref{fig:fv_SN80_0.3beam} for the
$S/N$= 80 results, and Fig.~\ref{fig:fv_SN30_0.3beam} for the $S/N$= 30
results.

Each line marks the trend in the normalised total fit difference if 
only the method
plotted on the abscissa is varied, and all other method variations are 
held constant. Sometimes the diamond symbols overlap, giving the impression that
there are fewer simulations in some of the panels. However, the number of
lines in each panel is the same and the variation of parameters is
distributed evenly over the various panels. The large spread in normalised
total fit difference of the different lines is primarily due to
binning the XV diagram along the major axis prior to fitting. In the 
plots showing
how the measurements varied with other method variations, the lines
with high measurement error are those which used a large major axis
binning interval. As expected, binning the XV diagram prior to fitting reduced 
the amount of major axis structure information, which dramatically increased 
the measurement uncertainty of all three fitted functions. The different
values of the binning interval have in some cases resulted in the outcomes being
distributeded in two somewhat distinct groups.

The plots show that smoothing the fits before incrementing, and
damping the model incrementation offer no improvement. For the high 
signal-to-noise dataset, the velocity dispersion measurements were 
improved by smoothing the velocity profile of each slice
by a small Gaussian prior to fitting.  The velocity dispersion 
measurements to the peak $S/N$= 80 dataset
were improved by $25\%$ (or on average $0.5-1.0$
\kms) with a Gaussian smoothing kernel with a FWHM of 2 velocity
channels, i.e.  slightly larger than the FWHM of noise which will
typically be 1-2 channels. However, Gaussian smoothing the velocity
axis caused no distinct improvement in the measurements to the low
signal-to-noise dataset. This is probably because Gaussian smoothing a
noisy profile distorts the shape slightly by incorporating the noise
into the smoothed profile shape.  
Consequently, we have not adopted
any additional smoothing or other variations discussed in this
section (Sect.~\ref{sec:ch4-disc-finetuning}) to the radial
decomposition method.

\subsection{Accuracy of the iterative radial decomposition method}
\label{sec:ch4-accuracy}

Convergence to the best kinematic model was usually achieved after three
iterations of the radial decomposition fitting process.  The average
measurement errors of each kinematic fit determined with the adopted
unchanged radial decomposition method are: $\Delta \sigma_{v,\HI}$=
0.7, 1.3 \kms\ (for peak $S/N$= 80 and peak $S/N$= 30,
respectively); $\Delta V$= 0.9, 1.4 \kms,  and 
$\Delta (\log \Sigma_{\HI)}$= 0.1, 0.15. The sum
of the residual XV diagram shows that the radial decomposition level was able
to recover the flux to a level of 0.2\% and 0.6\% of the total
flux of the observed XV diagram (for maps with peak $S/N$= 80 and $S/N$= 30,
respectively). This measurement accuracy is indicative only, as it constitutes
a mean error to measurements of  a galaxy with this particular
kinematics and structure. As mentioned earlier, fits to a XV map observed to
the same sensitivity level of a galaxy with a less steep intrinsic
surface density should have a lower measurement error.

\section{Summary}

Using our new iterative radial decomposition method, we show that it is
possible to accurately measure all three functions that describe the
planar kinematics and distribution of a galaxy: the radial \HI\ surface
density, the rotation curve and the \HI\ velocity dispersion. This
method is a significant improvement on previous methods to model
edge-on galaxies which approximated one or more of these kinematic
functions. Measurements of simulated galaxies show that a peak
signal-to-noise of $\gesim 30$ is needed to accurately measure all
three kinematic functions on a galaxy with a intrinsically steep
(exponential over all R) \HI\ surface density. However, reliable 
measurements would probably also be obtained with observations at a 
peak signal-to-noise
($S/N \gesim 20$) Furthermore, the method should also be successful 
on galaxies with an intrinsically
shallower surface density --- as is common in most disk galaxies due to 
depletion of \HI\ in the stellar disk.

In the next paper in this series, we present the derived \HI\ kinematics 
and radial surface density of our galaxies as measured with this method. 
Paper III also contains the vertical \HI\ flaring results, which is measured 
as a function of galactocentric radius using the deprojected \HI\ distribution 
and kinematics.

\begin{acknowledgements}

We are very grateful to Albert Bosma who contributed greatly to
initiating this project. He pointed out that \HI\ flaring studies are
best done on edge-on galaxies with low maximum rotational velocities,
and we used an unpublished Parkes \HI\ survey of edge-on galaxies by
Bosma and KCF when selecting our galaxies.  JCO thanks E. Athanassoula,
M.  Bureau, R.  Olling, A. Petric and J. van Gorkom for helpful
discussions.  JCO is grateful to B. Koribalski, R. Sault, L.
Staveley-Smith and R.  Wark for help and advice with data reduction and
analysis.
We thank the referee, J.M. van der Hulst, for
his careful and  thorough reading of the manuscripts of this series of papers and
his helpful and constructive remarks and suggestions.

\end{acknowledgements}

\bibliographystyle{aa}

\begin{thebibliography}{}
\bibitem[Binney \&\ Tremaine(1987)]{bt1987}
Binney, J. \& Tremaine, S., 1987, Galactic Dynamics (Princeton
University Press)
\bibitem[Bottema et al.(1986)]{bsvdk1986}
Bottema, R., Shostak, G.S. \&\ van der Kruit, P.C., 1986, \aap, 167, 34 
\bibitem[Boulanger \&\ Viallefond(1992)]{bv1992}
Boulanger, F. \&\ Viallefond, F., 1992, \aap, 266, 37
\bibitem[de Blok \&\ Bosma(2002)]{dbb2002}
de Blok, W.J.G. \&\ Bosma, A., 2002, \aap, 385, 816
\bibitem[Dickey et al.(1990)]{dhh1990}
Dickey, J.M., Hanson, M.M. \&\ Heloe, G., 1990, \apj, 352, 522
\bibitem[Fall \&\ Efstathiou(1980)]{fe1980}
Fall, S.M. \&\ Efstathiou, G., 1980, \mnras, 193, 189
\bibitem[Fraternali et al.(2002)]{fvmso2002}
Fraternali, F., van Moorsel, G., Sancisi, R. \&\ Oosterloo, T., 2002,
\aj, 123, 3124
\bibitem[Fraternali et al.(2005)]{foss05}
Fraternali, F., Oosterloo, T.A., Sancisi, R. and Swaters, R., 2005,
ASP Conf., 331.  239
\bibitem[Garc\'{\i}a-Ruiz et al.(2002)]{grsk2002}
Garc\'{\i}a-Ruiz, I., Sancisi, R. \&\ Kuijken, K.H., 2002, \aap, 394, 769
\bibitem[Kregel \&\ van der Kruit(2004)]{kvdk2004}
Kregel, M. \&\ van der Kruit, P.C., 2004, \mnras, 352, 787
\bibitem[Kregel et al.(2004)]{kvdkdb2004}
Kregel, M., van der Kruit, P.C. \&\ de Blok, W.J.G., 2004, \mnras, 352, 768
\bibitem[Lo et al.(1993)]{lsy1993}
Lo, K.Y., Sargent, W.L.W. \&\ Young, K., 1993, \aj, 106, 507
\bibitem[Mathewson et al.(1992)]{mfb1992}
Mathewson, D.S., Ford, V.L. \&\ Buchhorn, M., 1992, \apjs, 81, 413 
\bibitem[Nelder \&\ Mead(1965)]{nm1965}
Nelder, J.A. \& Mead, R., 1965, Computer J., 7, 308
\bibitem[Petric \&\ Rupen(2006)]{pr2006}
Petric, A.O. \&\ Rupen, M.P., 2007, \aj, 134, 1952 
\bibitem[O'Brien et al.(2010)]{ofk2008}
O'Brien, J.C., Freeman, K.C., van der Kruit, P.C. \&\ Bosma, A., 2010, \aap, submitted
(Paper I)
\bibitem[Olling(1996)]{olling1996a}
Olling, R.P., 1996, \aj, 112, 457
\bibitem[Rownd et al.(1994)]{rdh1994}
Rownd, B.K., Dickey, J.M. \&\ Helou, G., 1994, \aj, 108, 1638
\bibitem[Press et al.(1992)]{ptvf1992}
Press, W.H., Teukolsky, S.,A., Vetterling, W.T. \&\ Flannery, B.P., 1992,
Numerical Recipes in FORTRAN - The Art of Scientific Computing (Cambridge
University Press) 
\bibitem[Rubin et al.(1985)]{rbft1985}
Rubin, V.C., Burstein, D., Ford, W.K. \&\ Thonnard, N., 1985, \apj, 289, 81
\bibitem[Sancisi \&\ Allen(1979)]{sa1979}
Sancisi, R. \&\ Allen, R.J., 1979, \aap, 74, 73
\bibitem[Shane \&\ Bieger-Smith(1966)]{sbs1966}
Shane, W.W. \&\ Bieger-Smith, G.P., 1966, Bull. Astr. Inst. Neth., 18, 263
\bibitem[Shostak \&\ van der Kruit(1984)]{svdk1984}
Shostak, G.S. \&\ van der Kruit, P.C., 1984, \aap, 132, 20 
\bibitem[Sofue(1996)]{sofue1996}
Sofue, Y., 1986, \apj, 458, 120
\bibitem[Sofue et al.(2003)]{skno2003}
Sofue, Y., Koda, J. Nakamishi, H. \&\ Onodera, S., 2003, \pasj, 55, 59
\bibitem[Swaters(1999)]{swaters1999}
Swaters, R.S., 1999, Ph.D. Thesis, Groningen
\bibitem[Swaters et al.(1997)]{ssvdh1997}
Swaters, R.S., Sancisi, R. \&\ van der Hulst, J.M., 1997, \apj, 491, 140
\bibitem[Takamiya \&\ Sofue(2002)]{ts2002}
Takamiya, T. \&\ Sofue, Y., 2002, \apj, 576, L15
\bibitem[Uson \&\ Matthews(2003)]{um2003}
Uson, J.M. \&\ Matthews, L.D., 2003, \aj, 125, 2455
\bibitem[van der Kruit(1981)]{vdkruit1981}
van der Kruit, P.C., 1981, \aap, 99, 298
\bibitem[van der Kruit \&\ Shostak(1982)]{vdks1982c}
van der Kruit, P.C. \&\ Shostak, G.S., 1982, \aap, 105, 351
\bibitem[van der Kruit \&\ Shostak(1984)]{vdks1984}
van der Kruit, P.C. \&\ Shostak, G.S., 1984, \aap, 134, 258
\bibitem[Warmels(1988)]{warmels1988}
Warmels, R.H., 1988, \aaps, 72, 427
\end{thebibliography}

\end{document}